\documentclass[12pt]{article}
\usepackage{amsmath,amssymb}
\usepackage{comment}
\usepackage{bm}
\usepackage{color}
\usepackage{graphicx}
\usepackage[dvipdfmx]{epsfig}
\usepackage{epsfig}
\usepackage{braket}
\usepackage{color}
\usepackage{caption}
\usepackage{siunitx}
\usepackage[hidelinks=true]{hyperref}

\newcommand{\bea}{\begin{eqnarray}}
\newcommand{\eea}{\end{eqnarray}}

\DeclareMathAlphabet{\pazocal}{OMS}{zplm}{m}{n}

\usepackage{letltxmacro}
    \LetLtxMacro\oldl\l
    \DeclareRobustCommand{\l}{\ifmmode\lambda\else\oldl\fi}
    \LetLtxMacro\oldpr\partial
    \DeclareRobustCommand{\pr}{\ifmmode\partial\else\oldpr\fi}
    \LetLtxMacro\oldd\dagger
    \DeclareRobustCommand{\d}{\ifmmode\dagger\else\oldd\fi}
    \LetLtxMacro\oldno\noindent
    \DeclareRobustCommand{\no}{\ifmmode\noindent\else\oldno\fi}
    \LetLtxMacro\oldc\cos{}
    \DeclareRobustCommand{\c}{\ifmmode\cos{}\else\oldc\fi}
    \LetLtxMacro\oldc\sin{}
    \DeclareRobustCommand{\s}{\ifmmode\sin{}\else\olds\fi}
    \LetLtxMacro\oldth\theta
    \DeclareRobustCommand{\th}{\ifmmode\theta\else\oldth\fi}
    \LetLtxMacro\oldal\alpha
    \DeclareRobustCommand{\al}{\ifmmode\alpha\else\oldal\fi}
    
\begin{document}

\title{
\begin{flushright}
\ \\*[-80pt] 
\begin{minipage}{0.2\linewidth}
\normalsize
\end{minipage}
\end{flushright}
{\Large \bf 
Investigating Boson Sector in an Extended \\ Standard Model with $U(1)_D$ Symmetry  
\\*[20pt]}}
\author{ 
\centerline{
Apriadi Salim Adam$^{1}$\footnote{E-mail address: apriadi.salim.adam@brin.go.id},
Yunita Kristanti Andriani$^{1,2}$\footnote{E-mail address: ykandriani2871@gmail.com}, and
Eny Latifah$^{2}$\footnote{E-mail address: eny.latifah.fmipa@um.ac.id}
}
\\*[20pt]
\centerline{
\begin{minipage}{\linewidth}
\begin{center}
$^1${\it \normalsize
Research Center for Quantum Physics, National Research and Innovation Agency (BRIN)\\
South Tangerang 15314, Indonesia} \\*[5pt]
$^2${\it \normalsize
Department of Physics, Faculty of Mathematics and Natural Sciences, Universitas Negeri Malang, Jl. Semarang 5, Malang 65145,Indonesia} \\*[5pt]
\end{center}
\end{minipage}}
\\*[50pt]}
\date{
\centerline{\small \bf Abstract}
\begin{minipage}{0.9\linewidth}
\medskip 
\medskip 
\small
We have investigated the boson sector in an extended standard model (SM) with additional $U(1)_D$ symmetry. In the proposed model, the singlet scalar  and doublet scalar Higgs are added in addition to the SM-like scalar Higgs. These scalars are also coupled to the gauge boson fields. In this work, we calculate the masses of both gauge and scalar Higgs bosons. Their masses are obtained through spontaneous symmetry breaking using the Higgs fields with non-zero vacuum expectation values. We also study numerically the positivity conditions of the vacuum expectation value of the scalars. In particular, we perform scanning of the parameter space of the potential and study the obtained scalar mass dependence on the parameter of the model.
\end{minipage}  
}
\begin{titlepage}
\maketitle 
\thispagestyle{empty}
\end{titlepage}

\section{Introduction} \label{sec1}
The observation of a new boson by the ATLAS \cite{ATLAS:2012yve} and CMS \cite{CMS:2012qbp} experiments at the Large Hadron Collider in 2012 has solidified the Standard Model (SM) of particle physics. The observed spectrum of the new particle is precisely determined to be around 125 GeV \cite{ParticleDataGroup:2022pth} and confirms the Higgs mechanism for the electroweak gauge symmetry breaking \cite{Higgs:1964ia,Englert:1964et,Guralnik:1964eu,Higgs:1966ev,Kibble:1967sv}. Despite being a successful theory, the SM is well known to be incomplete as it could not explain several phenomena, such as the origin of the neutrino mass, baryon asymmetry of the Universe, hierarchy problems, dark matter, etc. These indicate a need to extend the SM. 

One of the ways to extend the SM is to introduce additional fields in the potential of the scalar field. Many different types of scalar potential have been proposed and these include the extensions of the SM-like additional singlet \cite{OConnell:2006rsp,Bowen:2007ia,Espinosa:2011ax,Pruna:2013bma}, two Higgs doublet model (2HDM) \cite{Haber:1978jt,Gunion:2002zf} (see also Ref.\cite{Branco:2011iw} for a review) and  minimally
supersymmetric SM Higgs (MSSM) \cite{Gunion:1984yn,Gunion:1986nh,Csaki:1996ks,Heinemeyer:2004ms,Draper:2016pys}. A general feature of all these potentials is the vacuum expectation value (VEV) of scalar fields that agrees with that of the SM. Concerning the models with more than one Higgs-like field, the physical Higgs either can be considered as an admixture of the fields \cite{Lopez-Val:2013yba,Chen:2014ask} or is already decoupled from the other Higgs \cite{Carena:2013ooa,BhupalDev:2014bir}. In the former case, the physical Higgs mass will constrain the mixing angle among the Higgs fields. There is another variant of the Higgs sector extension where one adds a singlet scalar and a (heavy) doublet scalar Higgs \cite{Gu:2006dc}. In this particular model, both extra singlet and doublet fields are charged under a new $U(1)_{X}$ gauge symmetry, while the SM fields are not. The heavy scalar doublet Higgs is integrated out; thus it can play a role in the mass generation of neutrinos through the effective dimension-5 operator. It is important to note that attempting to observe all of these additional Higgs fields so far has not shown any evidence yet.   

One characteristic of the extended SM with additional scalar fields is the possible new neutral gauge boson $Z'$ that may appear in the model. These additional scalars may generate the interaction between Higgs scalars and the gauge boson through their kinetic terms. In supersymmetric models, $Z'$ usually has a mass around the TeV scale \cite{Barger:1987xw,Espinosa:1997ji}. It also could be lighter than the electroweak scale if it has small interaction couplings with the SM particles. Many searches for the $Z'$ gauge bosons have been carried out and give experimental constraints on its parameters \cite{Langacker:1991pg,Liao:2017uzy,Papoulias:2017qdn,Bobovnikov:2018fwt,Cheung:2022oji,Hewett:1988fe,ALEPH:2005ab,Ramirez-Sanchez:2016ugz}. This extra gauge boson provides phenomenological implications as well as significant consequences for future collider physics and cosmology. These consequences have been reviewed comprehensively in \cite{Langacker:2008yv}.

Among various models that include extended Higgs sectors, this paper discusses a model with additional $U(1)_D\times U(1)_{B-L}$ gauge symmetry proposed in \cite{Dutta:2022knf}. This model attempts to simultaneously explain neutrino mass, dark matter, and the baryon asymmetry of the Universe by the TeV-scale physics without fine-tuning. In the model, they introduced two scalar doublets and two singlet scalars. After the scalars gain their VEV, the gauge boson and scalar Higgs will have mass. There is a notable lack of previous work \cite{Dutta:2022knf} where the mass of gauge bosons has not been explored. Thus to complement the proposed model, we will investigate the mass of the gauge boson and re-investigate the scalars Higgs sector in detail through this work. In particular, we perform a scanning of parameter space dependence of the potential, which will determine the mass scale of the scalar Higgs.

This paper is organized as follows. A description of the model is presented in Sec \ref{sec2}. We compute and investigate the mass generation of the scalar Higgs and gauge boson of the proposed model in Sec. \ref{sec3}. A numerical study for the positivity condition of the scalar VEVs is presented in Sec. \ref{sec4}. Section \ref{sec5} is devoted to our summary and an outlook.

\section{Description of The Model} \label{sec2}

To be self-contained, we briefly review an extended SM introduced in \cite{Dutta:2022knf}, which sets the framework for what we will analyze below, focusing on both scalar and gauge sectors. The model is based on the gauge group $SU(3)_c \times SU(2)_L \times U(1)_Y$ with additional $U(1)_D$ symmetry\footnote{In Ref. \cite{Dutta:2022knf}, another $U(1)_{B-L}$ gauge symmetry is also added to the gauge group of the model. The additional scalars ($X_i,\eta,\Phi',\Phi$) do not carry any hypercharge quantum number corresponding to $U(1)_{B-L}$ gauge symmetry (Please see Table 1 of Ref. \cite{Dutta:2022knf}). In this respect, they would not contribute to the mass of the gauge bosons through the covariant derivative. Based on this reason, we can omit this symmetry in our analysis.}.
The scalar fields content of the proposed model is defined in Table \ref{table:1}, where $X_{i}$ ($i=1,2$), $H$, and $\eta$ are the scalar doublets, while $\Phi$ and $\Phi'$ are the singlet scalars. In this table, we have also suppressed the generation indices for simplicity. The scalar doublet $H$ is defined as an SM-like scalar Higgs while $X_i$, being very heavy, does not engage in the low-energy process and has only implications in the leptogenesis scenario. On the other hand, the scalar $\eta$ is assumed to have zero vacuum expectation value (VEV) to preserve the $\mathbb{Z}_2$ symmetry and plays a role in facilitating the transfer of the lepton asymmetry from the dark sector to the visible sector. Under the $\mathbb{Z}_2$ symmetry, the scalar $\eta$ is CP-odd, while the others are CP-even. These additional scalar fields carry nontrivial charges under $U(1)_D$ symmetry, summarized in Table \ref{table:1}. 
\begin{table}[ht]
        \centering
        \begin{tabular}{c c c c}
            \hline \hline
            Scalar fields & $SU(2)_L$ & $U(1)_Y$ & $U(1)_D$ \\
            \hline
            $X_i$ & $2$ & $+1$ & $-1$ \\ 
            $\eta$ & $2$ & $+1$ & $1/2$ \\ 
            $H$ & $2$ & $+1$ & $0$ \\
            $\Phi'$ & $1$ & $0$ & $+1$ \\ 
            $\Phi$ & $1$ & $0$ & $0$ \\
            \hline \hline
        \end{tabular}
        \caption{The additional scalar fields with their quantum number with respect to the gauge group $SU(2)_L \times U(1)_Y\times U(1)_{D}$.}
        \label{table:1}
    \end{table}

 Now, let us write the scalar potential of the proposed model under the imposed symmetry. It is given by \cite{Dutta:2022knf},
\begin{equation} \label{eq.1}
    \begin{split}
        V &= \mu^2_1 (\eta^\d \eta) +\l_1 (\eta^\d \eta)^2 + \l_2 (\eta^\d \eta)(H^\d H) + \l_3[ (\eta^\d H)^2 + \mathrm{H.c.}] - \mu^2_2 H^\d H \\
        & \quad + \l_4 (H^\d H)^2 + \frac{1}{2} \mu^2_3 \Phi^2 + \frac{1}{3} \mu_4 \Phi^3 + \frac{1}{4} \l_5 \Phi^4 - \mu^2_5 (\Phi'^\d \Phi') + \l_6 (\Phi'^\d \Phi')^2 \\
        & \quad + \frac{\mu_6}{\sqrt{2}} \Phi (H^\d H) + \frac{\mu_7}{\sqrt{2}} \Phi (\Phi'^\d \Phi') + \frac{\l_7}{2} H^\d H \Phi^2 + \l_{8} H^\d H (\Phi'^\d \Phi') + \frac{\l_9}{2} \Phi^2 (\Phi'^\d \Phi').
    \end{split}
\end{equation}
The parameters $\mu_{i}$ ($i=1,..,7$) and $\lambda_{j}$ ($j=1,..,9$) can be chosen so that the scalar potential can lead to non-zero VEVs for certain scalars. These coupling constants are assumed to be real\footnote{In Ref.\cite{Dutta:2022knf}, the coupling $\l_{3}$ is set to be complex. In this study, we set this coupling as real for simplicity. This setting will not affect the analysis of the dimension-eight transfer operator, which appears at lower energy after the heavy scalar $\eta$ is integrated out (See Eq.(4) in Ref.\cite{Dutta:2022knf}).}. Note that the couplings $\mu_{i}$ have a mass dimension and represent the soft-breaking terms, while the other couplings are dimensionless. After the symmetry breaking occurs spontaneously, the scalars obtain their VEVs which are given by,
\begin{align} \label{eq.2}
    \Phi' = \frac{v_1}{\sqrt{2}};
    && H = \frac{1}{\sqrt{2}} \begin{pmatrix}
            0\\
            v_2	
    \end{pmatrix};
    && \Phi = v_3;
    && \eta = \frac{1}{\sqrt{2}} \begin{pmatrix}
            0\\
            v_4	
    \end{pmatrix},
\end{align}
where we may set the VEV of the scalar $\eta$ to be zero, $v_{4}=0$, as we have explained above. The above VEVs are assumed to have the following hierarchy, $v_1 > v_2 \gg v_3$. This choice of assumption for the VEVs agrees with the numerical analysis of the relevant phenomenology done in Ref. \cite{Dutta:2022knf}.

\section{Scalar Higgs and Gauge Boson Particles} \label{sec3}
In this section, we will focus on the mass generation for the scalar Higgs and the gauge boson at the low-energy level. After spontaneous symmetry breaking occurs, the scalars can gain their VEVs, and the mass of both scalar Higgs and gauge boson is obtained. 

\subsection{Gauge and Higgs Bosons in Electroweak Theory} \label{sec.3.1}
Before investigating the proposed model \cite{Dutta:2022knf}, we briefly review the mass generation of the gauge and Higgs bosons in the SM electroweak theory, which is based on the gauge group $SU(2)_{L} \times U(1)_{Y}$. 
We consider the Lagrangian of the SM for a doublet scalar Higgs as follows \cite{Halzen:1984mc},
\begin{equation} \label{eq.lsm}
    \pazocal{L} = \left( \pr_\mu \phi \right)^\d \left( \pr_\mu \phi \right) - \mu^2 \phi^\d \phi - \l \left( \phi^\d \phi \right)^2,
\end{equation}
where the doublet Higgs field is given by,
\begin{equation} 
    \phi =  \begin{pmatrix}
        \phi^{+} \\ \phi^{0}
    \end{pmatrix} .
\end{equation}
In the above Lagrangian, $\mu$ and $\lambda$ are the coupling constants. The condition that $\lambda >0$ and $\mu^{2} < 0$ are required for the scalar Higgs to be bounded from below. From Eq.\eqref{eq.lsm}, the potential of the scalar Higgs is written as, 
\begin{equation}\label{pot}
    \mathcal{V}(\phi) = \mu^2 \phi^\d \phi + \l \left( \phi^\d \phi \right)^2.
\end{equation}
The spontaneous symmetry breaking can occur if the scalar Higgs takes the following vacuum expectation value,
\begin{equation} \label{eq.vev}
    \langle \phi \rangle = \frac{1}{\sqrt{2}} \begin{pmatrix}
        0 \\ v
    \end{pmatrix} 
\end{equation}
with $v=\sqrt{-\mu^{2}/\lambda}$.

As a consequence of the spontaneous symmetry breaking of the Higgs field, the fermion fields can get their masses via Yukawa interactions except for the neutrino (the SM does not accommodate the right-handed neutrino)\footnote{The detailed derivation for the mass generation of the fermion fields lies beyond the scope of the present study. For one who wishes to study their mass generation, please see Ref.\cite{Halzen:1984mc} for instance.}. Besides that, the gauge boson $W$ and $Z$ can also obtain their masses. The Lagrangian of the gauge boson mass terms is given by \cite{Halzen:1984mc},
\begin{equation}\label{bostera}
    \pazocal{L}_{b} = \left| \left( i\pr_\mu - g \bm{T} \cdot \bm{W_\mu} - g' \frac{Y}{2} B_\mu \right) \phi \right|^2,
\end{equation}
where $\bm{T}=\tau_{i}/2$ ($i=1,2,3$) and $\tau$ is the Pauli matrix. The coupling constants $g$ and $g'$ represent the couplings of the gauge group $SU(2)_L$ and $U(1)_Y$, respectively. $Y$ is the hypercharge that corresponds to the charge operator $Q_{EW}$ and $T^{3}$, namely,
\begin{equation}
    Q_{EW} = T^3 +\frac{Y}{2}.
\end{equation}

Substituting the VEV of scalar Higgs Eq.\eqref{eq.vev} in Eq.\eqref{bostera}, one obtains 
\begin{align}\label{Lb}
     \pazocal{L}_{b} &= \frac{1}{8} \left|{\begin{pmatrix}  {g W^3_\mu + g' B_\mu} & g(W^1_\mu -i W^2_\mu) \\ g(W^1_\mu+i W^2_\mu) & {- g W^3_\mu + g' B_\mu} \end{pmatrix}} {\begin{pmatrix} 0 \\ v \end{pmatrix}} \right|^2 \nonumber \\
        &= \frac{1}{8} v^2 g^2 \left[ (W^1_\mu)^2 + (W^2_\mu)^2 \right] + \frac{1}{8} v^2 (- g W^3_\mu + g' B_\mu)^2 \nonumber \\
        &= \frac{1}{4} v^2 g^2 {W^{+}_\mu}{W^{-\mu}} + \frac{1}{8} v^2 \begin{pmatrix}
            W^3_\mu & B_\mu
        \end{pmatrix}
        \begin{pmatrix}
            g^2 & -gg' \\ -gg' & g'^2
        \end{pmatrix}
        \begin{pmatrix}
            W^{3\mu} \\ B^\mu
        \end{pmatrix}
\end{align}
where $W^{\pm}_{\mu}$ is defined as,
\begin{equation}
    W^{\pm}_{\mu} = \frac{(W^{1}_{\mu} \mp iW^{2}_{\mu})}{\sqrt{2}}.
\end{equation}
The first term of Eq.\eqref{Lb} leads directly to the mass of the charged gauge boson given by,
\begin{equation}\label{MW}
    M_{W^{\pm}} = \frac{1}{2} vg.
\end{equation}
To diagonalize the $2\times 2$ mass mixing matrix in the second term of Eq.\eqref{Lb}, one introduces the following new basis for the gauge boson, 
\begin{align} \label{eq.za}
    Z_\mu = \frac{g W^3_\mu  - g' B_\mu}{\sqrt{g^2 + g'^2}} , \quad & \quad
    A_\mu = \frac{g' W^3_\mu  + g B_\mu}{\sqrt{g^2 + g'^2}}.
\end{align}
By doing so, one can obtain the mass of a neutral gauge boson $Z_\mu$ and a photon $A_{\mu}$ as follows,
\begin{align}
    M_Z = \frac{1}{2} v \sqrt{g^2 + g'^2}, \qquad M_A = 0.
\end{align}

Next, to generate the mass of the scalar Higgs particle, one can expand the field around the classical minimum as \cite{Halzen:1984mc},
\begin{equation} \label{expd}
    \phi (x) = \frac{1}{\sqrt{2}} \begin{pmatrix}
        0 \\ v + h(x)
    \end{pmatrix}
\end{equation}
where $h(x)$ denotes the quantum fluctuations about this minimum. Substituting Eq.\eqref{expd} into Eq.\eqref{pot} and considering only the squared term of the field $h(x)^{2}$ lead to the mass of the scalar Higgs which is given by,
\begin{equation}
    M_h = \sqrt{2 \l v^2}.
\end{equation}

\subsection{Gauge Boson Sector in a Model with \texorpdfstring{$U(1)_D$}{U(1)D} Symmetry} \label{sec.3.2}
To determine the mass of the gauge boson particles in the present model, we consider the following relevant Lagrangian analog in the electroweak standard model as follows, 
\begin{equation} \label{eq.11}
    \pazocal{L}_{\mathrm{Boson}} = \left| D_\mu \Phi' \right|^2 + \left| D_\mu H \right|^2 + \left| D_\mu \Phi \right|^2
\end{equation}
where $D_\mu$ is the covariant derivative written as,
\begin{equation}
    D_\mu = \pr_\mu - ig \bm{T} \cdot \bm{W_\mu} - ig' \frac{Y}{2} B_\mu - ig'' X C_\mu
\end{equation}
with $X$ denoting the electroweak hypercharge operator and $g ''$ representing the couplings of the gauge group $U(1)_D$. The corresponding charge operator $Q_{D}$ of the gauge group in this model satisfies the following, 
\begin{equation}
    Q_{D} = T^{3} + \frac{Y}{2} + X.
\end{equation}

Next, by using the corresponding quantum number of the scalar fields in Table \ref{table:1} and using the VEVs of the scalars in Eq.\eqref{eq.2}, the Lagrangian in Eq.\eqref{eq.11} becomes, 
\begin{align}
     \label{eq.lb}
        \pazocal{L}_{\mathrm{Boson}} &= \left| \left( - ig \bm{T} \cdot \bm{W_\mu} - ig' \frac{Y}{2} B_\mu - ig'' X \bm{C_\mu} \right) \left( \frac{v_1}{\sqrt{2}} \right) \right|^2 \nonumber \\
        & \quad + \left| \left( - ig \bm{T} \cdot \bm{W_\mu} - ig' \frac{Y}{2} B_\mu - ig'' X \bm{C_\mu} \right) \begin{pmatrix}
            0\\
            \frac{v_2}{\sqrt{2}}	
        \end{pmatrix} \right|^2 \nonumber \\
        & \quad + \left| \left( - ig \bm{T} \cdot \bm{W_\mu} - ig' \frac{Y}{2} B_\mu - ig'' X \bm{C_\mu} \right) \left( v_3 \right) \right|^2 \nonumber \\
        &= \frac{1}{4} v_2^2 g^2 {W^{+}_\mu}{W^{-\mu}} + \frac{1}{2} \begin{pmatrix}
        W^3_\mu & B_\mu & C_\mu
        \end{pmatrix} M_{\mathrm{WB}} \begin{pmatrix}
            W^{3\mu} \\ B^\mu \\ C^\mu
        \end{pmatrix}
\end{align}
where $M_{\mathrm{WB}}$ is the mass mixing matrix of the bosons $W^3_\mu$, $B_\mu$, and $C_\mu$
\begin{equation} \label{Mbw}
    M_{\mathrm{WB}} = \frac{1}{4} \begin{pmatrix}
        v_2^2 g^2 & -v_2^2 gg' & 0 \\
        -v_2^2 gg' & v_2^2 g'^2 & 0 \\
        0 & 0 & 4 v_1^2 g''^2
    \end{pmatrix}. 
\end{equation}
The first term of Eq.\eqref{eq.lb} leads us directly to the same form for the mass of the charged gauge boson, $W^{\pm}$, given in Eq.\eqref{MW} setting $v$ with $v_{2}$. We note that in Eq.\eqref{eq.lb}, there will be a new neutral gauge boson different from the usual electroweak theory. 

To obtain the mass of the neutral gauge bosons and the photon field, one needs to diagonalize the mass mixing matrix $M_\mathrm{WB}$ in Eq.\eqref{eq.lb} (or in Eq.\eqref{Mbw}). The diagonalization is done by introducing a matrix $S$ which acts to the bosons $W^3_\mu$, $B_\mu$, and $C_\mu$ basis and forming the following new basis of the gauge boson fields, 
\begin{equation}
  \left(\begin{array}{c}
    Z_{\mu}\\
    A_{\mu}\\
    Z'_{\mu}
  \end{array}\right) = \left(\begin{array}{ccc}
    \cos \theta_W & - \sin \theta_W & 0\\
    \sin \theta_W & \cos \theta_W & 0\\
    0 & 0 & 1
  \end{array}\right) \left(\begin{array}{c}
    W^3_{\mu}\\
    B_{\mu}\\
    C_{\mu}
  \end{array}\right)
\end{equation}
where we have used the definition ${g'}/{g} = \tan{\th_W}$ with $\th_W$ being the weak mixing angle between gauge boson $W^3_\mu$ and $B_\mu$. Taking all these procedures, we obtain the masses of the expected gauge boson $Z$, a new gauge boson $Z'$ as follows,
\begin{equation}
    \qquad M_{Z} = \frac{1}{2} v_2 \sqrt{g^2 + g'^2}, \qquad M_{Z'} = v_1 g'', \label{gaugeboson}
\end{equation}
and the photon field remains massless. From the above equation, we note that the coupling $g''$ and the VEV of the scalar Higgs $\Phi'$ will determine the mass scale of the new gauge boson $Z'$.

\subsection{Scalar Higgs Sector in a Model with \texorpdfstring{$U(1)_D$}{U(1)D} Symmetry} \label{sec.3.3}
Below we reinvestigate the mass of the scalar Higgs boson in the present model \cite{Dutta:2022knf}. For this purpose, we first derive the VEVs of the fields $\Phi$, $\Phi'$, and $H$ written in terms of the parameters of the potential Eq.\eqref{pot}. These values are obtained by taking the derivative of the potential Eq.\eqref{eq.1} to its VEV $v_i$ as follows,
\begin{equation}
    \frac{\pr V}{\pr v_i} = 0,
\end{equation}
and we obtain
\begin{equation}
    \begin{split}
        v_1 \left( - \mu^2_{5} + \l_{6} v^2_1 + \frac{\mu_7}{\sqrt{2}} v_3 + \frac{\l_8}{2} v^2_2 + {\frac{\l_{9}}{2} v^2_3} \right) = 0 \\
        v_2 \left( - \mu^2_2 + \l_4 v^2_2 + \frac{\mu_6}{\sqrt{2}} v_3 + {\frac{\l_{7}}{2} v^2_3} + \frac{\l_{8}}{2} v^2_1 \right) = 0 \\
        v_3 \left( \mu^2_3 + {\mu_4 v_3} + {\l_5 v^2_3} + \frac{\l_7}{2} v^2_2 + \frac{\l_{9}}{2} v^2_1 \right) + \frac{\mu_6}{2 \sqrt{2}} v^2_2 + \frac{\mu_7}{2 \sqrt{2}} v^2_1 = 0.
   \end{split}
\end{equation}
Note that in the above equation, we have assumed $\mu_4 \ll \mu_3$ and $\l_5 \ll 1$ so that we can drop the terms proportional to $\pazocal{O}(v^2_3)$ and $\pazocal{O}(v^3_3)$. Thus solving the above equations simultaneously leads us to the following forms of the VEVs,
\begin{align} \label{eq:19}
     v_1 &= \sqrt{\frac{\mu^2_{5} - \frac{\mu_7}{\sqrt{2}} v_3 - \frac{\l_{8}}{2} v^2_2}{\l_{6}}}, \\
     \label{eq.41}
     v_2 &= \sqrt{\frac{\mu^2_2 - \frac{\mu_6}{\sqrt{2}} v_3 - \frac{\l_{8}}{2} v^2_1}{\l_4}}, \\
     \label{eq.42}
     v_3 &= \frac{-\left(\mu_6 v^2_2 + \mu_7 v^2_1 \right)}{2 \sqrt{2} \left(\mu^2_3 + \frac{\l_{7}}{2} v^2_2 + \frac{\l_{9}}{2} v^2_1 \right)}.
\end{align}

\begin{figure}[tb]
\begin{center}
\begin{tabular}{cc}
\includegraphics[width=4.6cm]{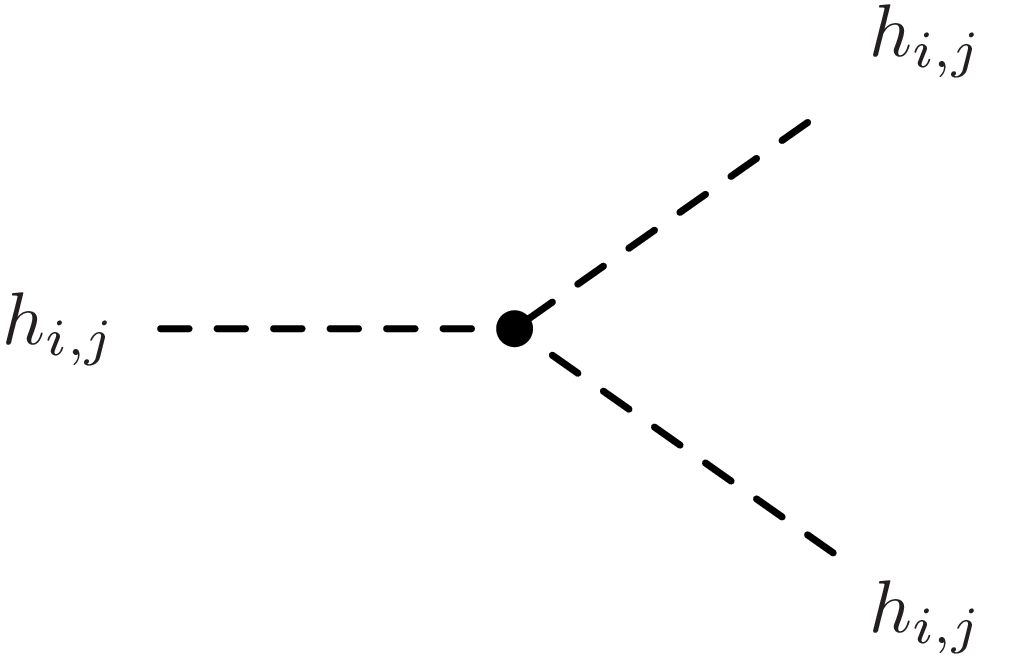} & 
\includegraphics[width=4.5cm]{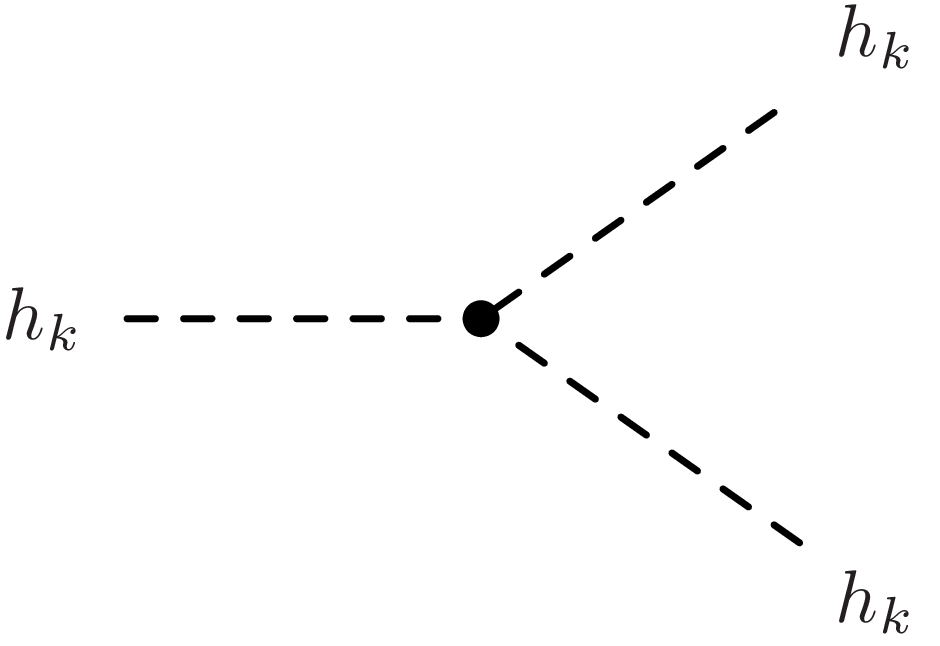}\\
\includegraphics[width=4.6cm]{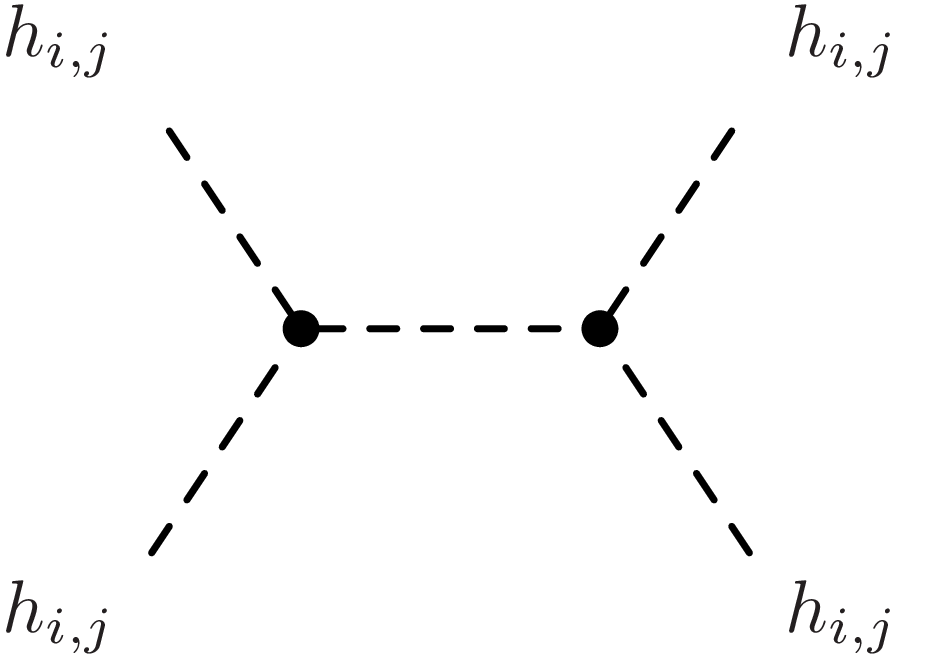} &
\includegraphics[width=4.5cm]{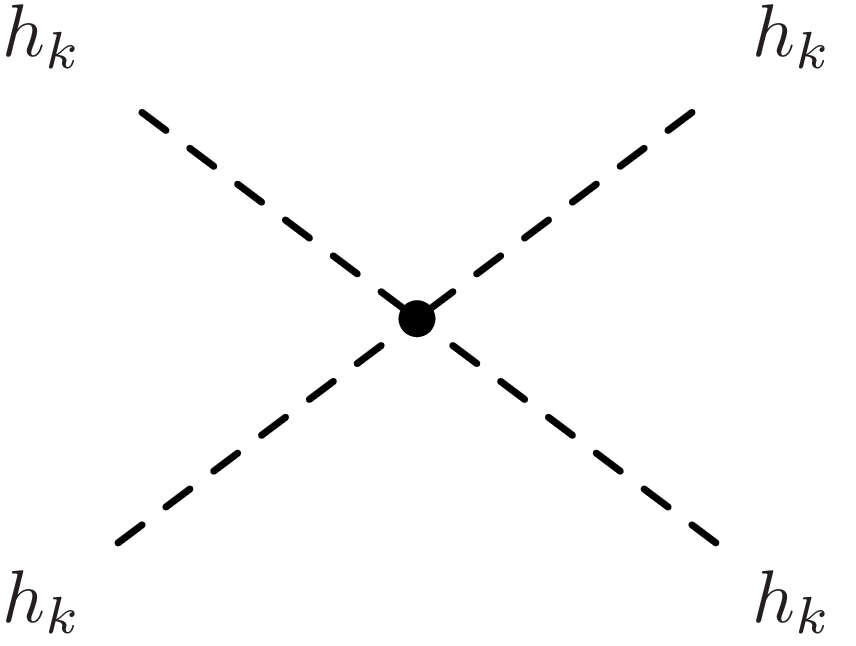}\\
\end{tabular}
\end{center}
\caption{The diagram of the tree-level decay and scattering processes, respectively (left panel) and the cubic and quartic self-interactions of the scalar Higgs 
 (right panel).}\label{fig:1}
\end{figure}

Next, we consider the mass of the scalar Higgs. Analog as in the electroweak theory, we expand the fields in Eq.\eqref{eq.2} with their corresponding fluctuation fields as 
\begin{align} \label{eq.2.1}
    \Phi' = \frac{v_1 + h_1}{\sqrt{2}};
    && H = \frac{1}{\sqrt{2}} \begin{pmatrix}
            0\\
            v_2 + h_2	
    \end{pmatrix};
    && \Phi = v_3 + h_3;
    && \eta = \frac{1}{\sqrt{2}} \begin{pmatrix}
            0\\
            0 + h_4	
    \end{pmatrix}
\end{align}
We drop the space-time $x$ dependence in the fluctuation fields for simplicity. Substituting Eq.\eqref{eq.2.1} into potential in Eq.\eqref{eq.1}, we obtain
\begin{equation}
    \begin{split} \label{eq.39}
        V =& \left( - \mu^2_5 v_1 + \l_{6} v_1^3 + \frac{\mu_7}{\sqrt{2}} v_1 v_3 + \frac{\l_{8}}{2} v_1 v_2^2 \right) h_1 + \left( - \mu^2_2 v_2 + \l_4 v_2^3 +\frac{\mu_6}{\sqrt{2}} v_2 v_3 + \frac{\l_{8}}{2} v_1^2 v_2 \right) h_2 \\
        &  + \left( \mu^2_3 v_3 + \frac{\mu_6}{2\sqrt{2}}v_2^2 + \frac{\mu_7}{2\sqrt{2}} v_1^2 + \frac{\l_{7}}{2} v_2^2 v_3 \right) h_3 + \frac{1}{2} \begin{pmatrix} h_1 & h_2 & h_3 \end{pmatrix}
        M^2_h
        \begin{pmatrix} h_1 \\ h_2 \\ h_3 \end{pmatrix} + \l_6 v_1 h^3_1 \\
        & + \l_4 v_2 h^3_2 + \left( \frac{1}{3} \mu_4 + \l_5 v_3 \right) h^3_3 + \frac{\l_8}{2} v_2 h^2_1 h_2 + \left( \frac{\mu_7}{2\sqrt{2}} + \frac{\l_9}{2} v_3 \right) h^2_1 h_3 + \frac{\l_9}{2} v_3 h_1 h^2_3 \\
        &  + \frac{\l_7}{2} v_2 h_2 h^2_3 + \left( \frac{\mu_6}{2\sqrt{2}} + \frac{\l_7}{2} v_3 \right) h^2_2 h_3 + \frac{\l_8}{4} h^2_1 h^2_2 + \frac{\l_9}{4} h^2_1 h^2_3 + \frac{\l_7}{4} h^2_2 h^2_3 + \frac{\l_6}{4} h^4_1 + \l_4 h^4_2 \\
        & + \frac{1}{4} \l_5 h^4_3 + \mathrm{constant}.
    \end{split}
\end{equation}
The first three terms proportional to linear $h_{i}$ do not correspond to any physical phenomena. The second term of the second line in the above equation is identified as the squared mass matrix $M^2_h$ of the corresponding scalar Higgs fields $\Phi'$, $H$, and $\Phi$ given by,
\begin{equation} \label{eq.6}
        M^2_h =
        \begin{pmatrix}
            2 \l_{6} v^2_1 & \frac{1}{2} \l_{8} v_2v_1 & \frac{1}{2} \left( \frac{\mu_7}{\sqrt{2}} v_1 + \l_{9} v_3v_1 \right) \\
            \frac{1}{2} \l_{8} v_2v_1 & 2 \l_4 v_2^2 & \frac{1}{2} \left( \frac{\mu_6}{\sqrt{2}}v_2 + \l_{7} v_3v_2 \right) \\
            \frac{1}{2} \left( \frac{\mu_7}{\sqrt{2}} v_1 + \l_{9} v_3v_1 \right) & \frac{1}{2} \left( \frac{\mu_6}{\sqrt{2}}v_2 + \l_{7} v_3v_2 \right) & \mu^2_3 + \frac{\l_{7}}{2} v_2^2 + \frac{\l_{9}}{2} v_1^2 
       \end{pmatrix}.
 \end{equation}
In Eq.\eqref{eq.39}, the terms $h_{i}^{2}h_{j}$ (or $h_{i}h_{j}^{2}$) and $h_{i}^{2}h_{j}^{2}$ ($i,j=1,2,3$) are related to the decay and scattering processes of the scalar Higgs, respectively, while the terms $h_{k}^{3}$ and $h_{k}^{4}$ ($k=1,2,3$) represent the interaction of the scalar Higgs with itself. We plot the tree-level diagram of those interactions in Fig.\ref{fig:1}.
Note that, since the scalar Higgs $h_{4}$ has already decoupled from other scalar Higgs at a high-energy level \cite{Dutta:2022knf}, it leads directly to its effective mass given by,
\begin{equation}
    {m}^2_{H_4} = \mu^2_1 + \frac{v^2_2}{2} \left( \l_2 + 2 \l_3 \right).
\end{equation}

Now to obtain the expected scalar Higgs $h_i$ ($i=1,2,3$) in mass basis, the mass matrix in Eq.\eqref{eq.6} must be diagonalized. It can be done by introducing a unitary matrix $M$ up to order $\sin \alpha$ and a rotation matrix $R$ with an $h_1$-$h_2$ mixing angle $\beta$ as follows, 
\begin{align} \label{eq.7}
    M = \begin{pmatrix}
        \c{\al} & 0 & \s{\al} \\
        0 & \c{\al} & \s{\al} \\
        - \s{\al} & - \s{\al} & \c{\al} \\
    \end{pmatrix}
    && \mathrm{and} &&
    R = \begin{pmatrix}
        \c{\beta} & - \s{\beta} & 0 \\
        \s{\beta} & \c{\beta} & 0 \\
        0 & 0 & 1 \\
    \end{pmatrix}.
\end{align}
Applying these matrices to the squared mass mixing matrix in Eq.\eqref{eq.6} allows us to write the new mass basis with diagonal form as,
\begin{align} 
        m^2_{H_1} &= 2 \l_6 v^2_1 + \frac{1}{2} \l_8 v_1 v_2 s_{2\beta} \nonumber \\
        & \quad - \frac{1}{4} s_{2\al} \left( \mu_7 v_1 \sqrt{2} + \frac{1}{\sqrt{2}} (\mu_6 v_2 + \mu_7 v_1) s_{2\beta} + \l_7 v_2 v_3 s_{2\beta} + \l_9 v_1 v_3 (2+s_{2\beta}) \right) \label{eq.8} \\
        m^2_{H_2} &= 2 \l_4 v^2_2 - \frac{1}{2} \l_8 v_1 v_2 s_{2\beta} \nonumber \\
        & \quad + \frac{1}{8} s_{2\al} \left( - 2 v_2 (2 \l_7 v_3 + \mu_6 \sqrt{2}) + \left( 2 v_3 (\l_7 v_2 + \l_9 v_1) + \sqrt{2} (\mu_6 v_2 + \mu_7 v_1) \right) s_{2\beta} \right) \label{eq.8b} \\
        m^2_{H_3} &= \frac{1}{2}  \left( 2 \mu^2_3 + \l_{7} v_2^2 + \l_{9} v_1^2 + \left( v_3 (\l_{7}v_2 + \l_{9}v_1) + \frac{1}{\sqrt{2}} (\mu_6 v_2 + \mu_7 v_1) \right) s_{2\al} \right) \label{eq.8c}
\end{align}
where we have used short-hand notations: $s_{2\beta} = \s{2\beta}$ and $s_{2\al} = \s{2\al}$. The detailed derivation of the above expression is given in Appendix \ref{app.A}. We can also obtain the relation of the mixing angle $\alpha$ and $\beta$ in terms of the potential parameters explicitly as follows, 
\begin{equation} \label{eq.tanAl}
    \tan{\al} = \frac{\frac{1}{2} \left( \frac{\mu_7}{\sqrt{2}} v_1 + \l_{9} v_1 v_3 \right)}{ \mu^2_3 + \frac{\l_{7}}{2} v_2^2 + \frac{\l_{9}}{2} v_1^2 - 2 \l_{6} v_1^2 - \frac{1}{2} \l_{8} v_1 v_2} ,
\end{equation}
\begin{equation} \label{eq.tanB}
    \tan{2\beta} = \frac{\l_{8} v_1 v_2 - \frac{1}{2} s_{2\al} \left(  \l_{7} v_2 v_3 + \frac{\mu_1 v_2}{\sqrt{2}}+ \l_{9} v_1 v_3 + \frac{\mu_7 v_1}{\sqrt{2}} \right)}{-2 \l_4 v_2^2 + 2 \l_{6} v_1^2 + \frac{1}{2} s_{2\al}  \left( \l_{7} v_2 v_3 + \frac{\mu_6 v_2}{\sqrt{2}} - \left( \l_9 v_1 v_3 + \frac{\mu_7 v_1}{\sqrt{2}} \right) \right)} .
\end{equation}

\section{Numerical Study} \label{sec4}
\begin{figure}[tb]
    \centering
    \includegraphics[width=9cm]{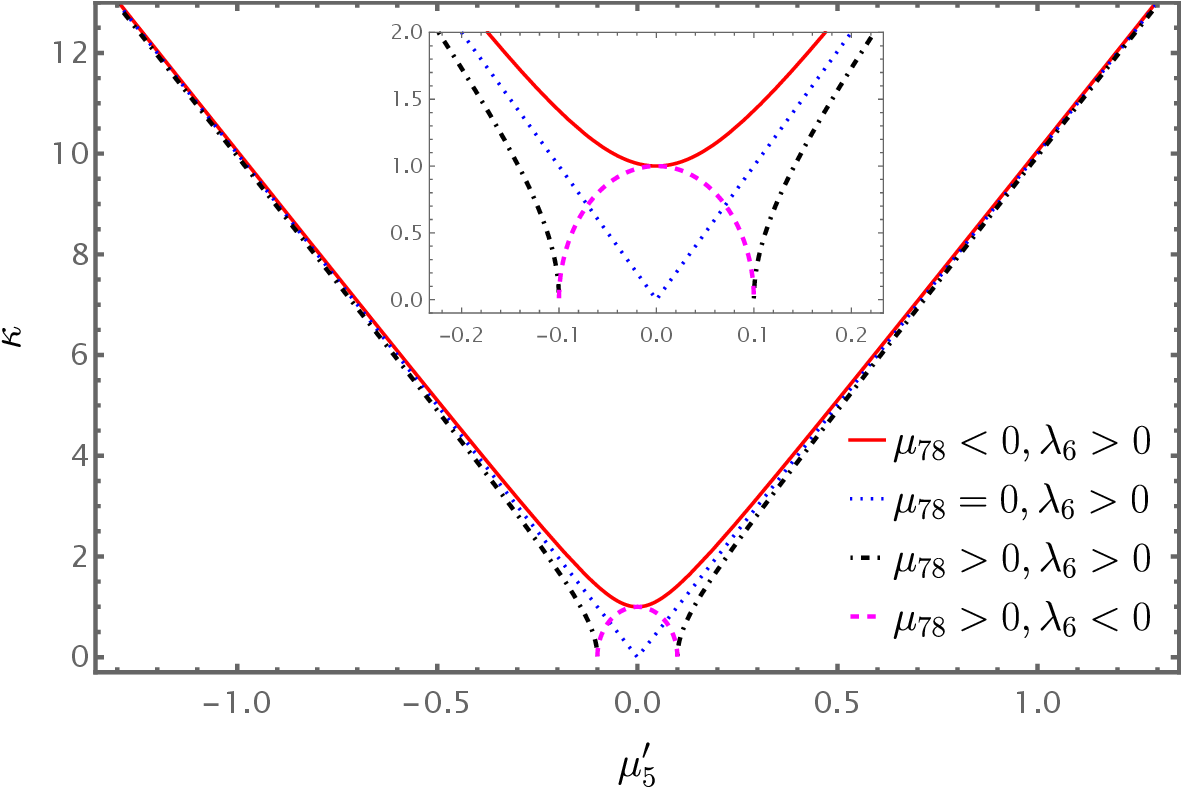}
    \caption{We show the parameter $\kappa$ as a function of the dimensionless coupling $\mu'_{5}$ for four possible conditions. The solid red, dotted blue, and dot-dashed black lines show the cases for $\l_6>0$ with three different values of $\mu_{78}<0$, $\mu_{78}=0$, and $\mu_{78}>0$, respectively, while the dashed magenta line shows the case of $\l_6<0$ and $\mu_{78}>0$. We fix a set of the parameter values as $\mu_{78}=(-10^{-2},0,10^{-2})$ and $\lambda_6=(- 10^{-2},10^{-2})$ for an illustration.}
    \label{fig:2}
\end{figure}

\begin{figure}[!tb]
    \centering
    \begin{tabular}{cc}
        \includegraphics[width=6cm]{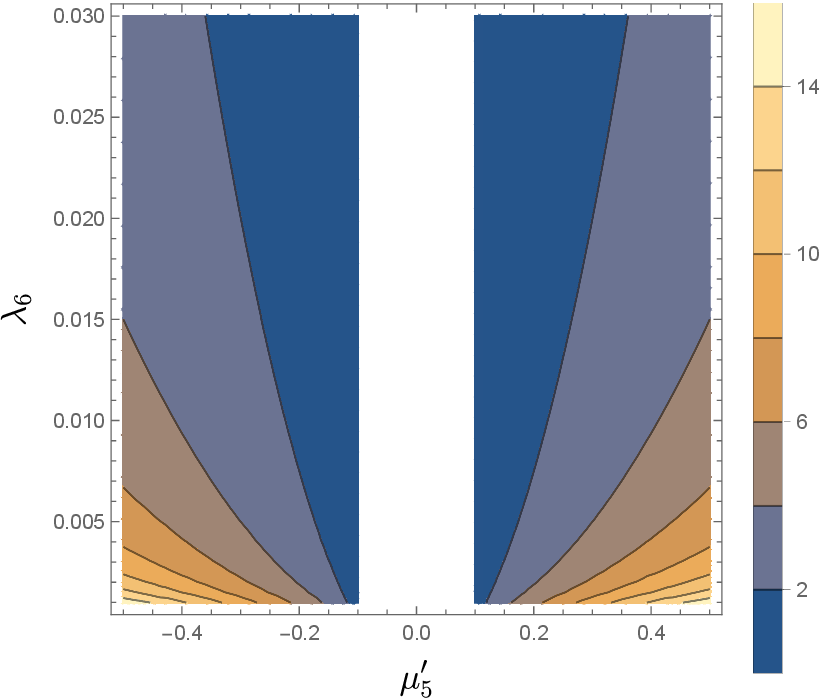} & \includegraphics[width=6cm]{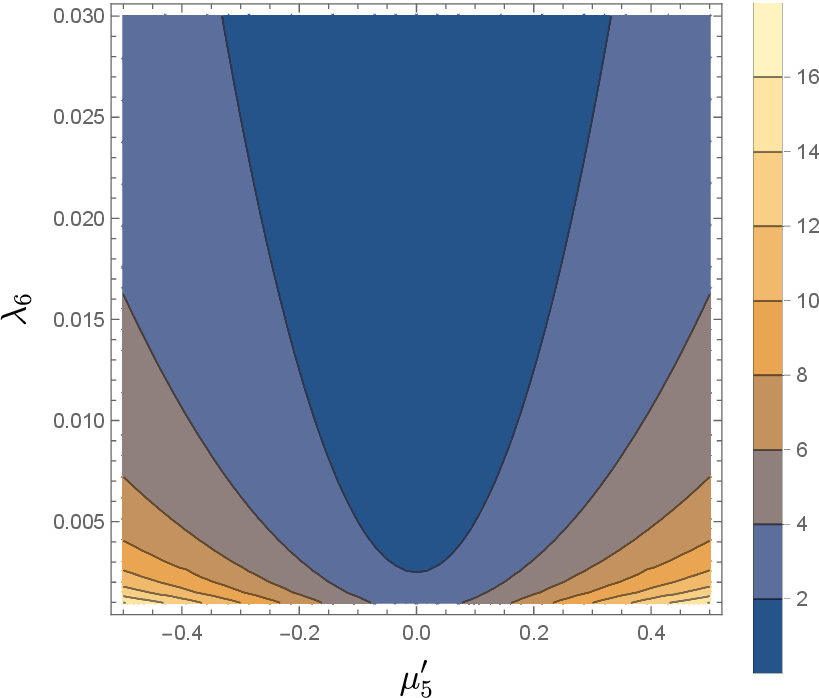} \\
        \includegraphics[width=6cm]{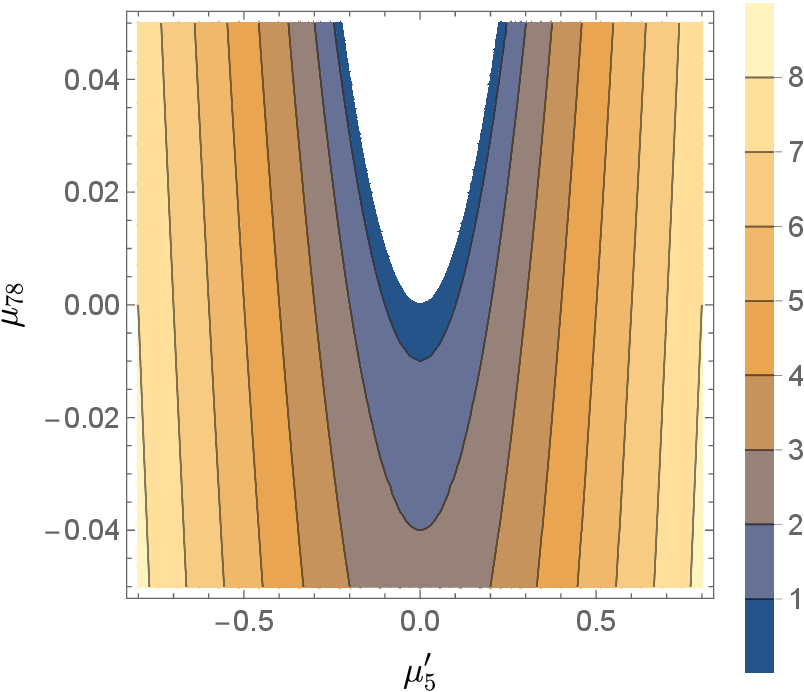} & \includegraphics[width=5.8cm]{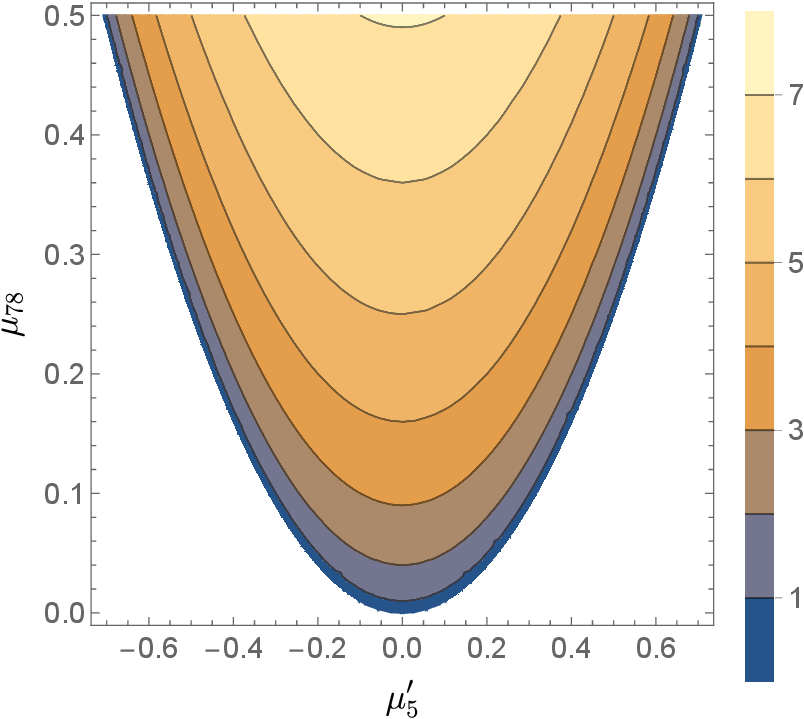} \\
        \multicolumn{2}{c}{\includegraphics[width=6.3cm]{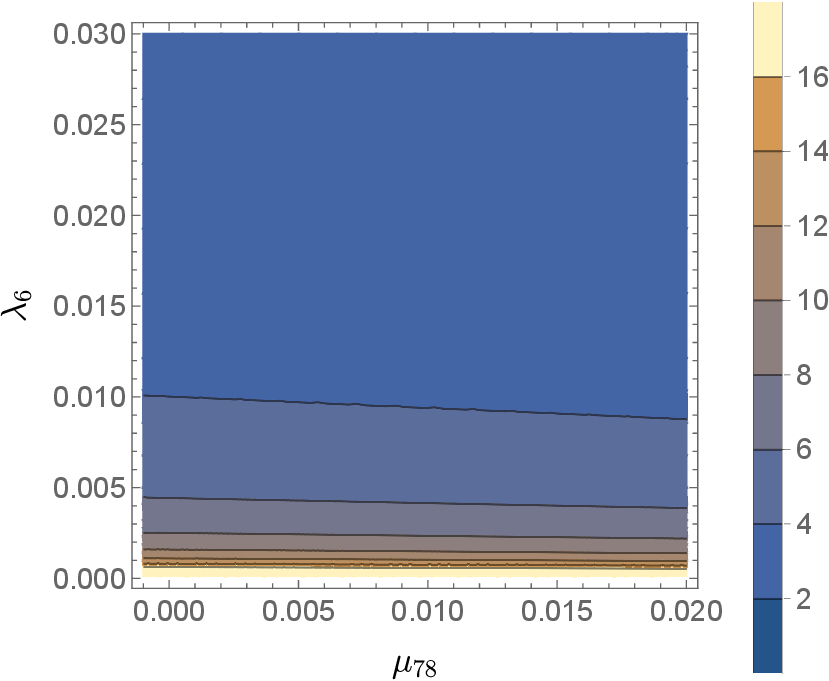}}
    \end{tabular}
    \caption{Contour plots show the configuration of $\kappa$. The top panel shows the configuration in the $\mu'_5 - \l_6$ plane with $\mu_{78}=10^{-2}$ (left) and $\mu_{78}=-10^{-2}$ (right). The middle panel shows the configuration in the $\mu'_5 - \mu_{78}$ plane for $\l_6=10^{-2}$ (left) and $\l_6=-10^{-2}$ (right). Meanwhile, the bottom panel shows the configuration in the $\mu_{78} - \l_6$ plane with $\mu'_5 = 0.4$. The non-allowed region of $\kappa$ is given in a white area in the above plots.}
    \label{fig:9}
\end{figure}

\begin{figure}[tb]
    \centering
    \includegraphics[width=9cm]{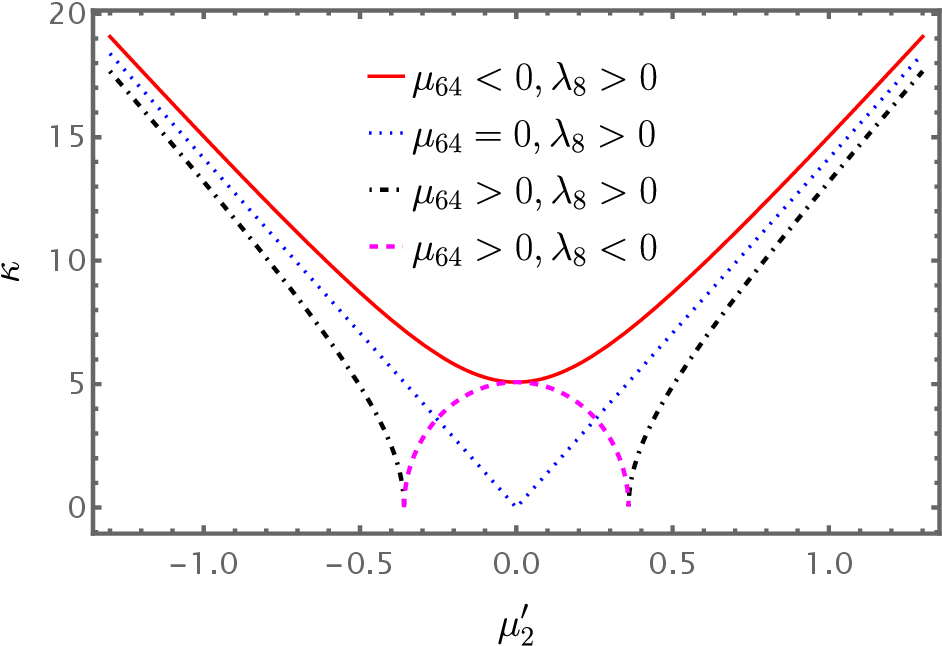}
    \caption{We show the parameter $\kappa$ as a function of the dimensionless coupling $\mu'_{2}$ for four possible conditions. The solid red, dotted blue, and dot-dashed black lines show the cases for $\l_8>0$ with three different values $\mu_{64}<0$, $\mu_{64}=0$, and $\mu_{64}>0$, respectively, while the dashed magenta line shows the case for $\l_8<0$ and $\mu_{64}>0$. We fix a set of parameters as $\mu_{64}=(-0.129,0,0.129)$ and $\l_8=(-10^{-2},10^{-2})$ for an illustration.}
    \label{fig:3}
\end{figure}

\begin{figure}[!htb]
    \centering
    \begin{tabular}{cc}
        \includegraphics[width=6.1cm]{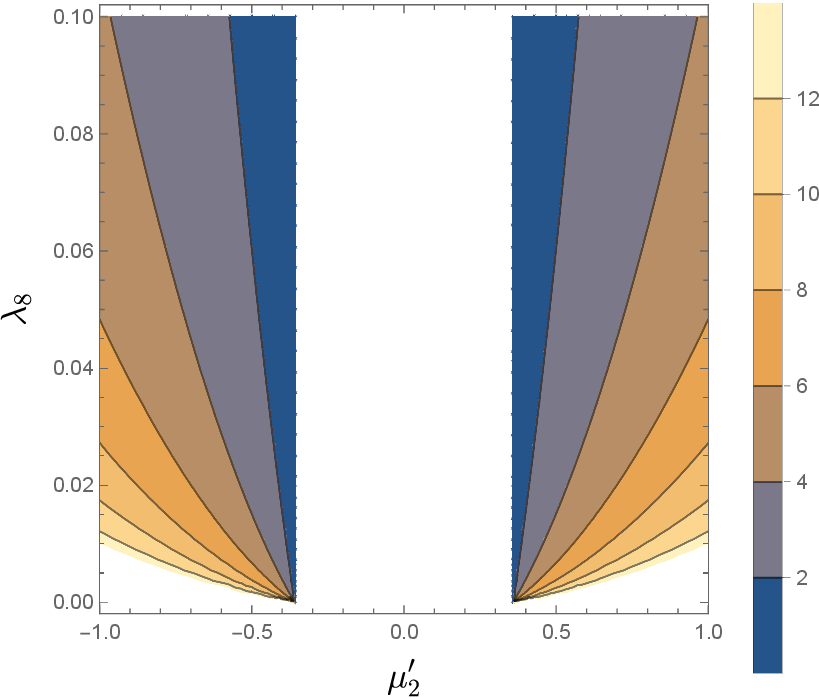} & \includegraphics[width=6.2cm]{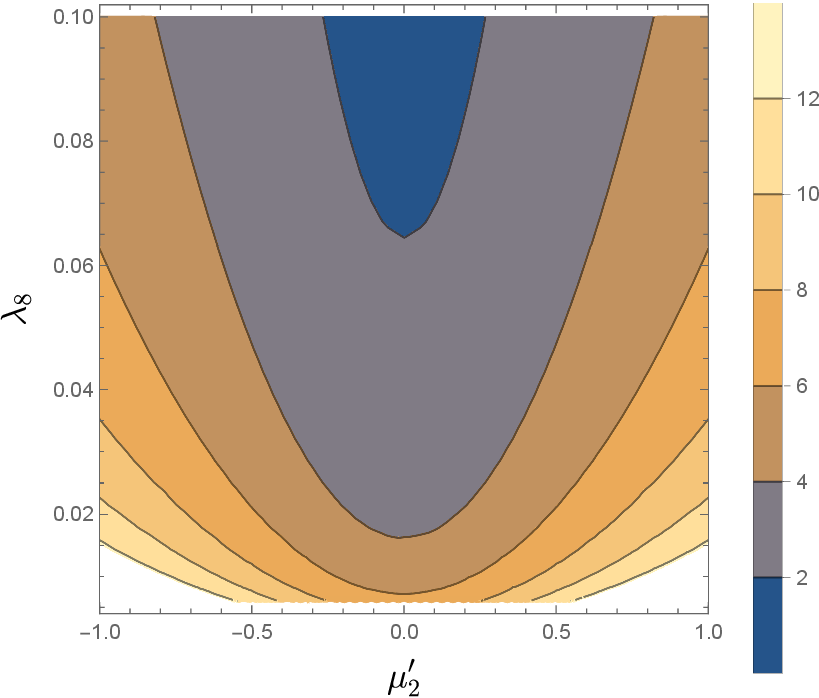} \\
        \includegraphics[width=6cm]{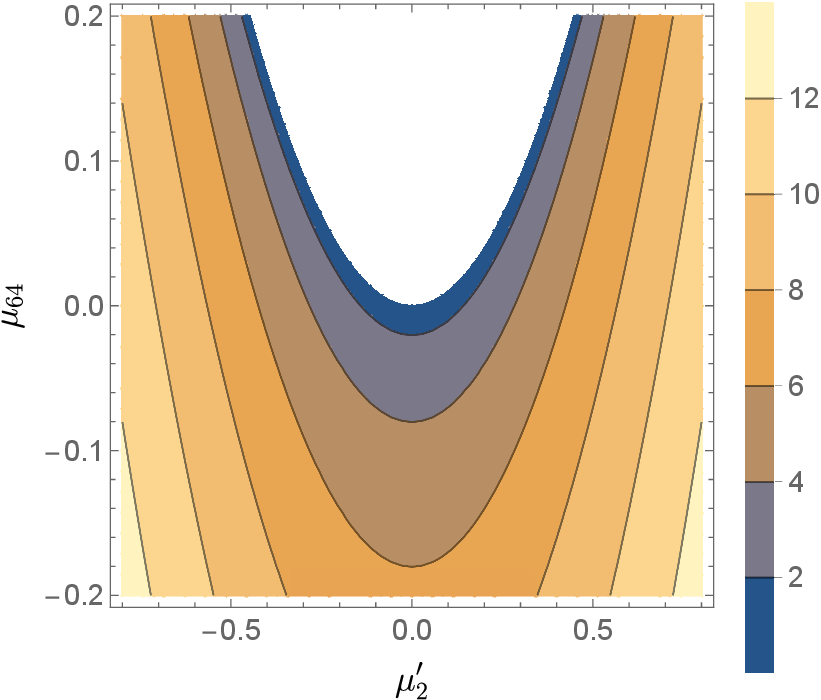} & \includegraphics[width=6cm]{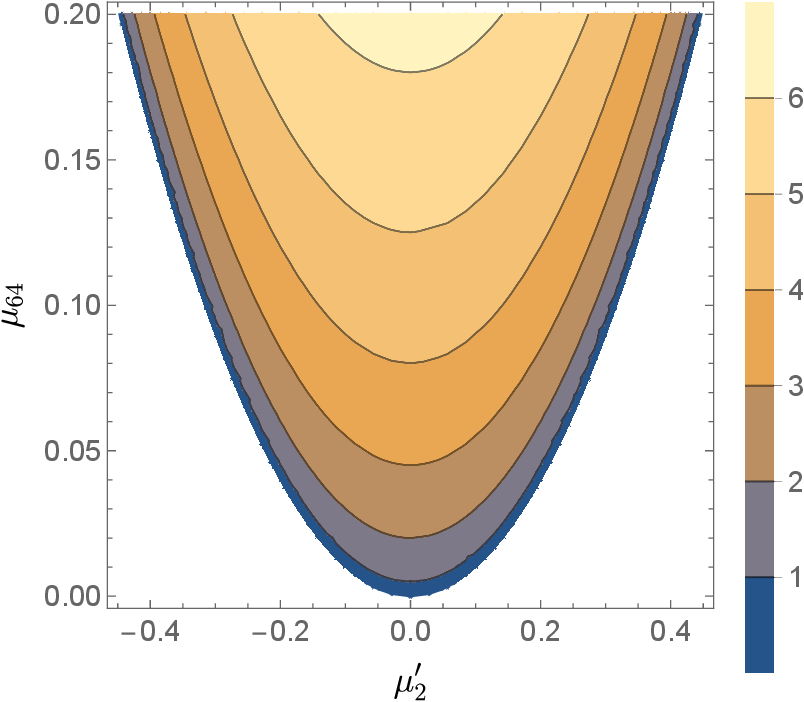} \\
        \multicolumn{2}{c}{\includegraphics[width=6cm]{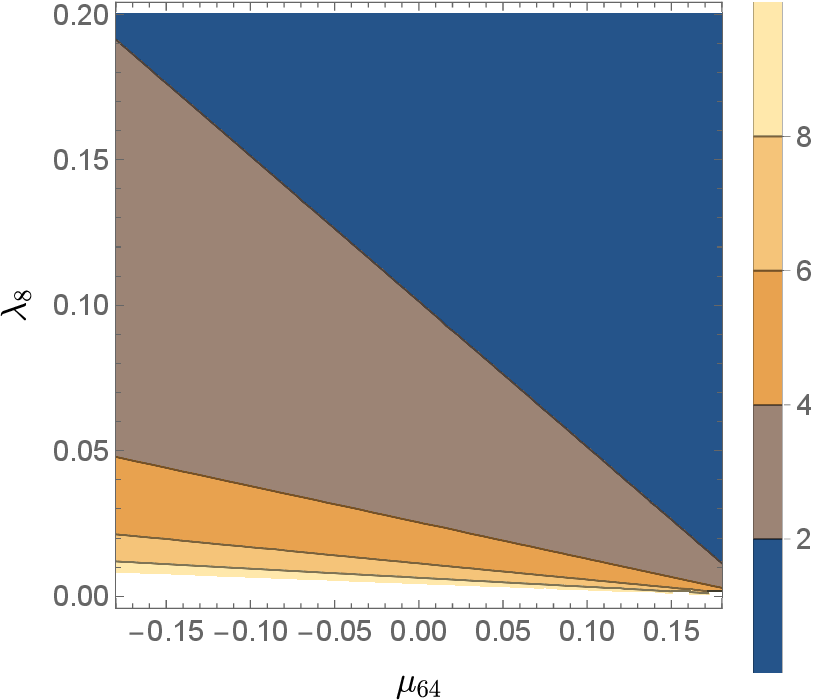}}
    \end{tabular}
    \caption{Contour plots show the configuration of $\kappa$. The top panel shows the configuration in the $\mu'_2 - \l_8$ plane with $\mu_{64}=0.129$ (left) and $\mu_{64}=-0.129$ (right). The middle panel shows the configuration in the $\mu'_2 - \mu_{64}$ plane for $\l_8 = 10^{-2}$ (left) and $\l_8=-10^{-2}$ (right). Meanwhile, the bottom panel shows the configuration in the $\mu_{64} - \l_8$ plane with $\mu'_2 =0.45$. The non-allowed region of $\kappa$ is given in a white area from the above plots.}
    \label{fig:10}
\end{figure}

\begin{figure}[tb]
    \centering
    \includegraphics[width=8.6cm]{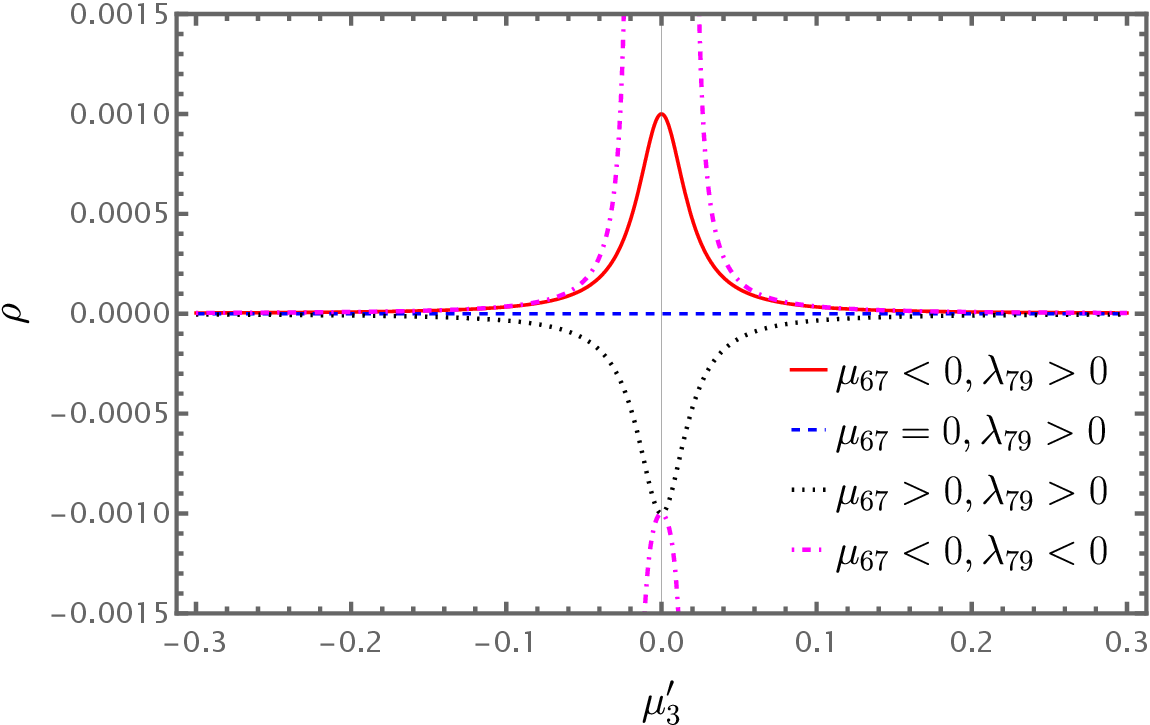}
    \caption{We show the parameter $\rho$ as a function of the dimensionless coupling $\mu'_{3}$. The solid red, dashed blue, and dotted black lines show the cases for $\l_{79}>0$ with three different values of $\mu_{67}<0$, $\mu_{67}=0$, and $\mu_{67}>0$, respectively, whereas the dot-dashed magenta line shows the case for $\l_{79}<0$ and $\mu_{67}<0$. Note that the dashed blue line is overlapped with the horizontal axis. We fix a set of parameter values as $\mu_{67}=(-10^{-6},0,10^{-6})$ and $\l_{79}=(-10^{-3},10^{-3})$ for an illustration.}
    \label{fig:4}
\end{figure}

\begin{figure}[!tb]
    \centering
    \begin{tabular}{cc}
        \includegraphics[width=6.5cm]{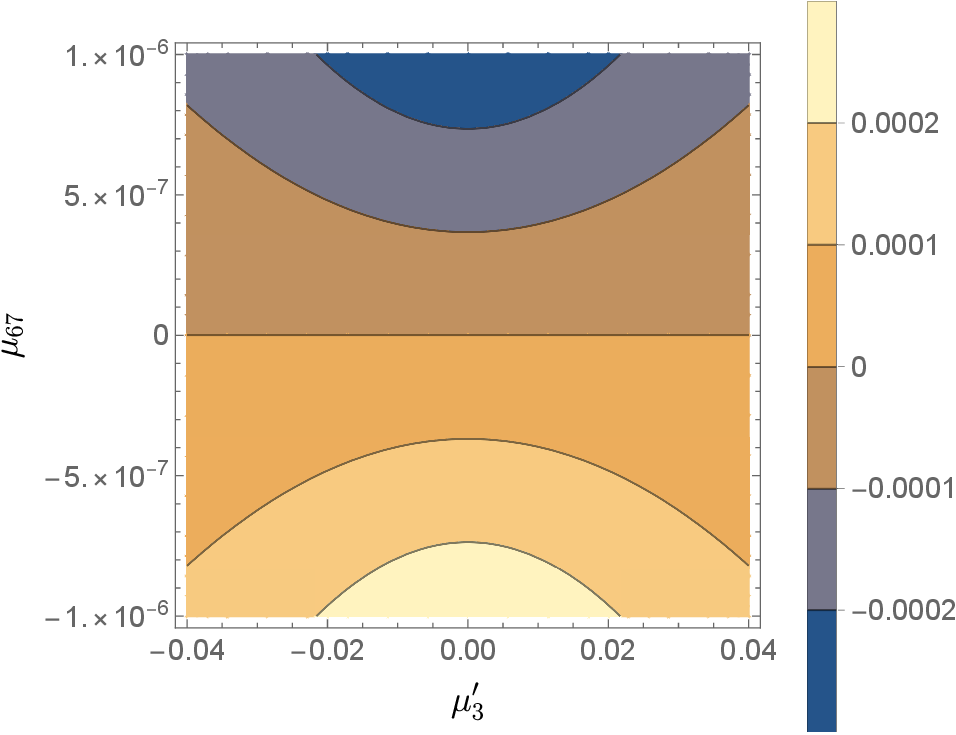} & \includegraphics[width=6.5cm]{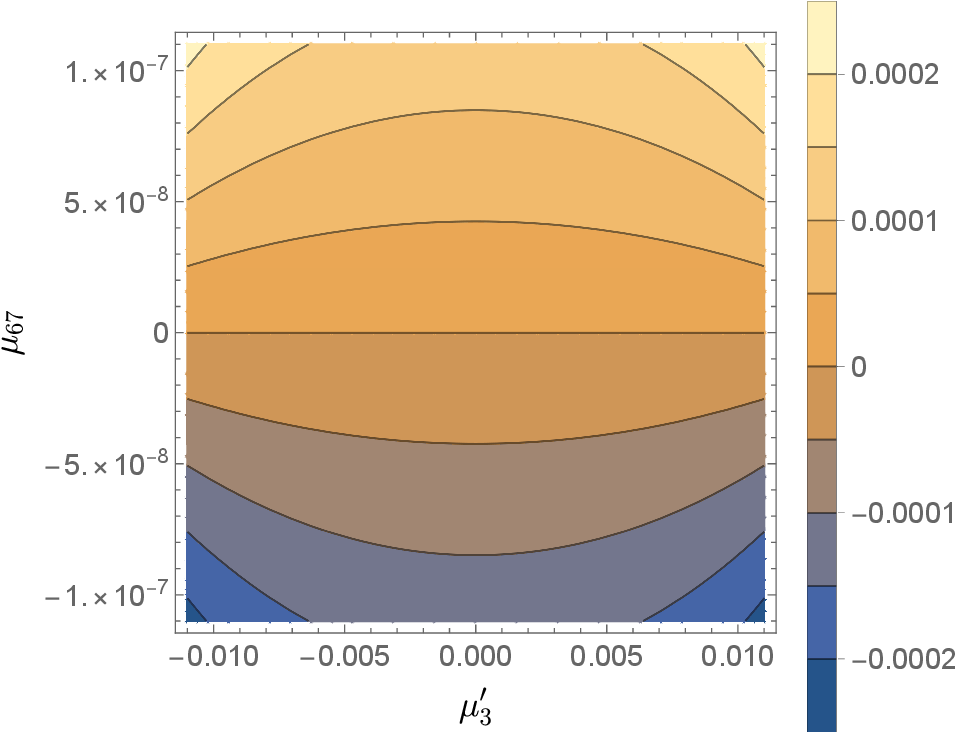} \\
        \includegraphics[width=6.5cm]{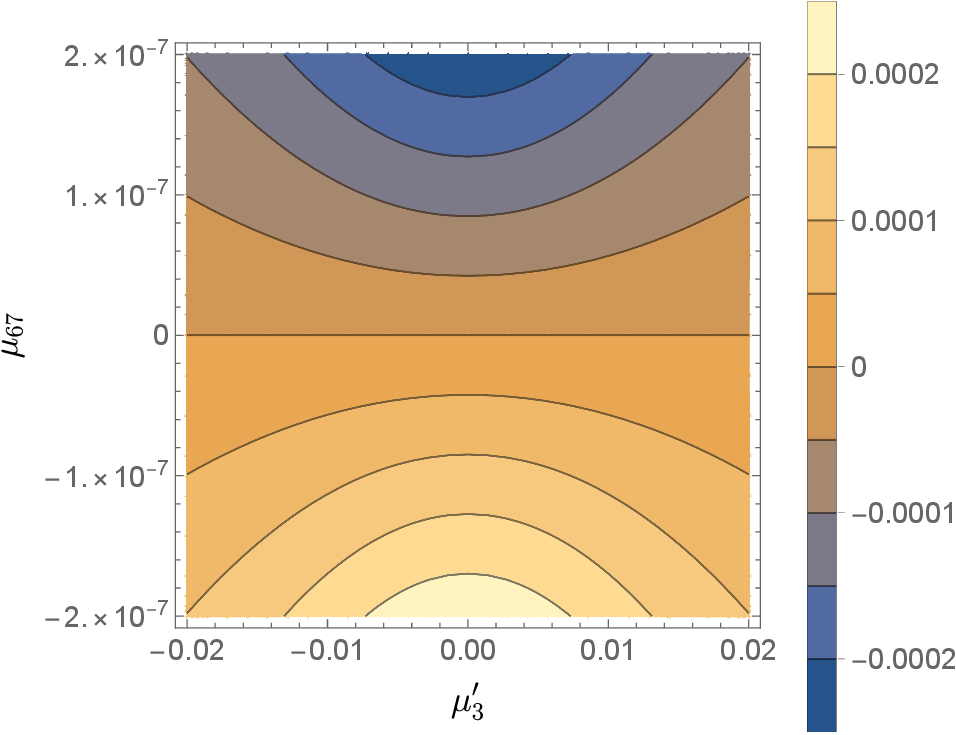} & \includegraphics[width=6.5cm]{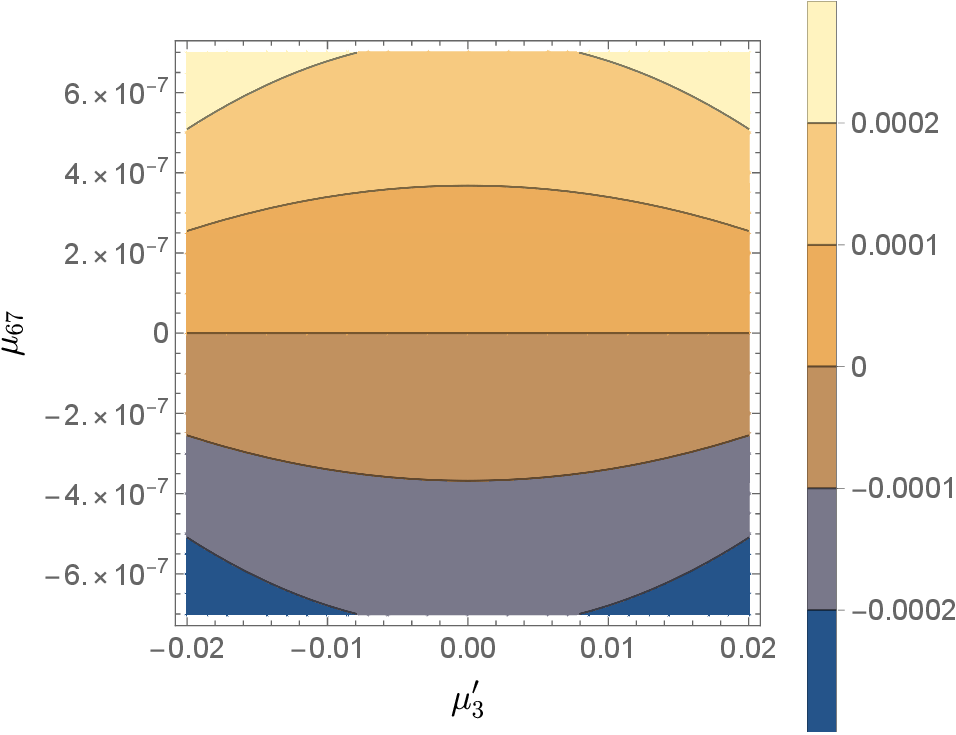}
    \end{tabular}
    \caption{Contour plots show the configuration of $\rho$ in the $\mu'_3 - \mu_{67}$ plane. The upper panel shows the case for $\l_7>0,\l_9>0$ (left) and $\l_7>0,\l_9<0$ (right). The lower panel shows the case for $\l_7<0,\l_9>0$ (left) and $\l_7, \l_9<0$ (right). We fix values for $\kappa = 4$, $\l_7 = 10^{-3}$, and $\l_9 = 10^{-4}$.}
    \label{fig:11}
\end{figure}

\begin{figure}[!tb]
    \centering
    \begin{tabular}{cc}
        \includegraphics[width=6cm]{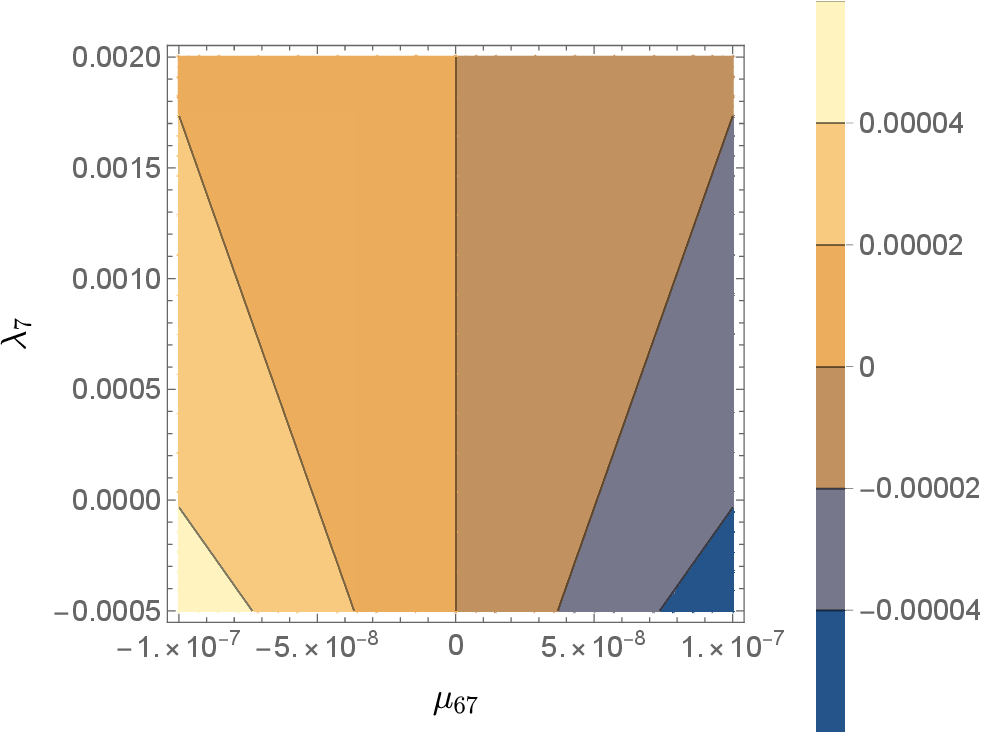} & \includegraphics[width=5.8cm]{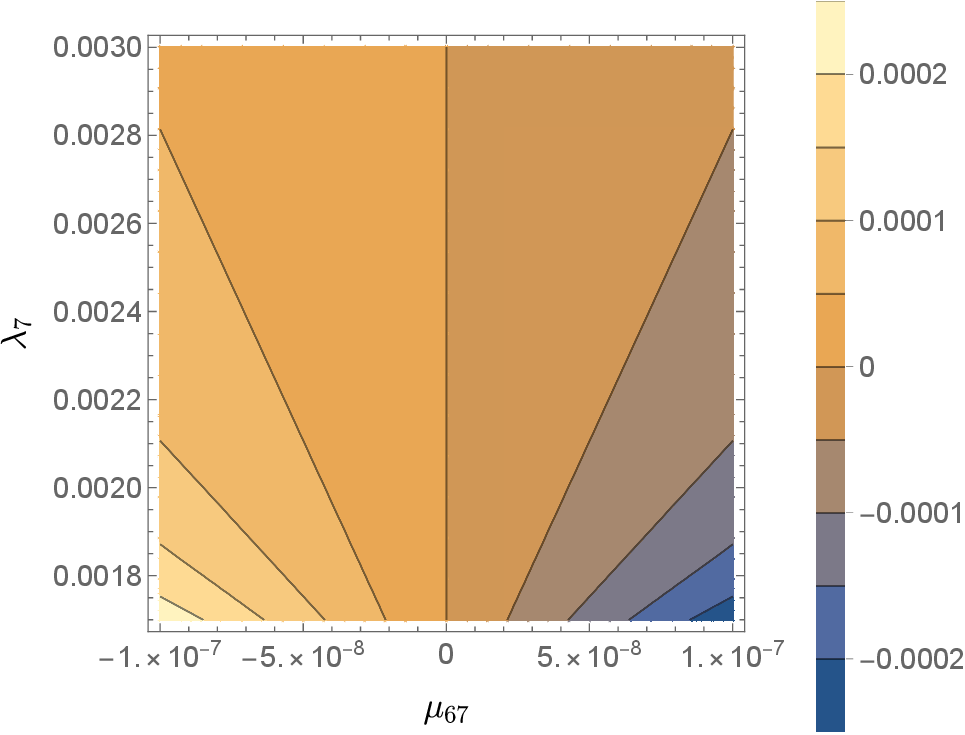} \\
        \includegraphics[width=6.2cm]{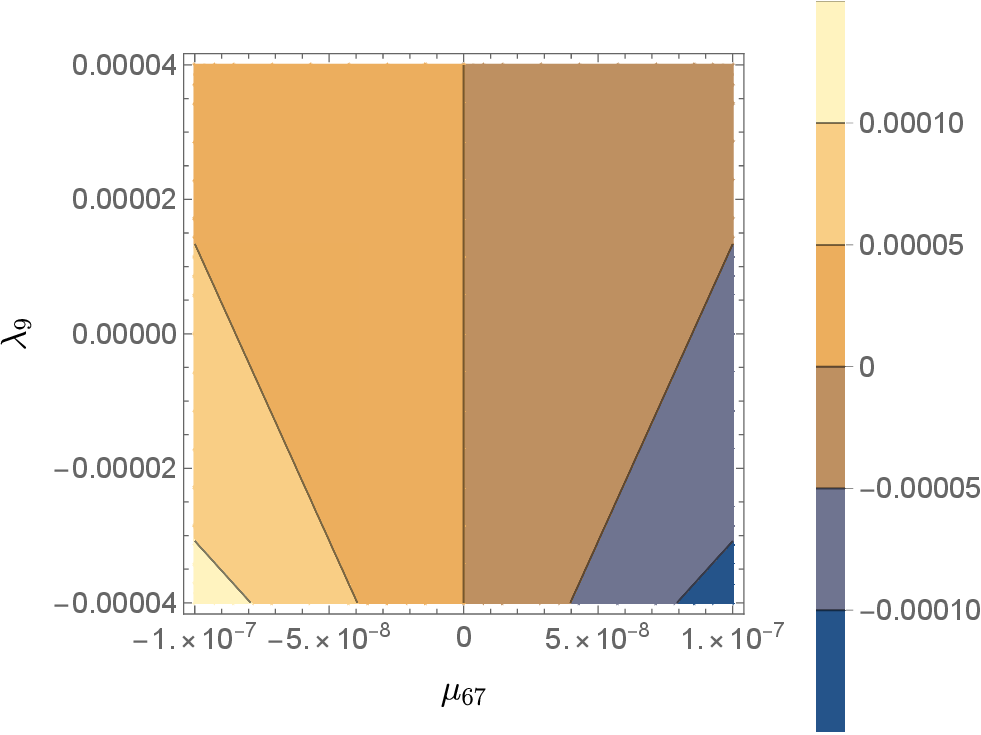} & \includegraphics[width=6.2cm]{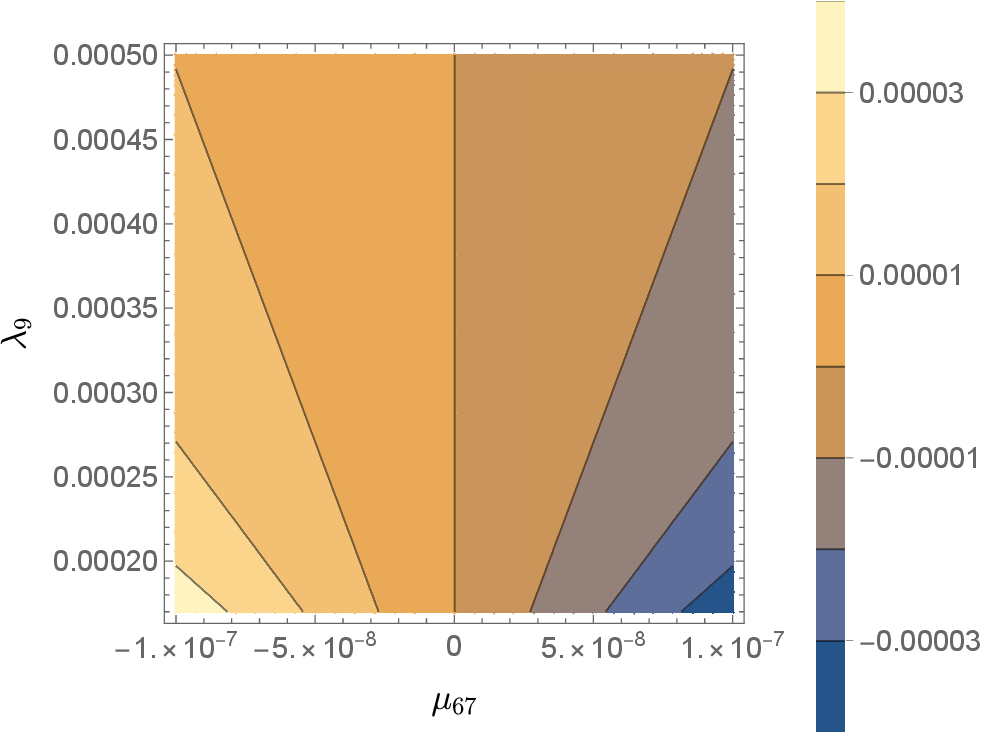} \\
        \includegraphics[width=6cm]{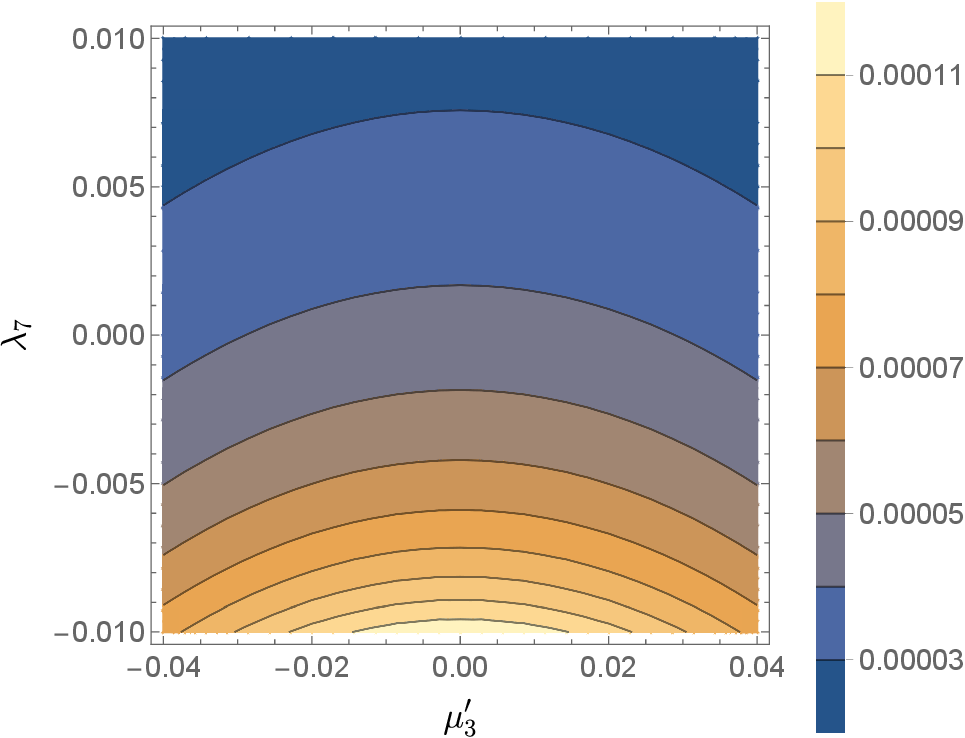} & \includegraphics[width=6cm]{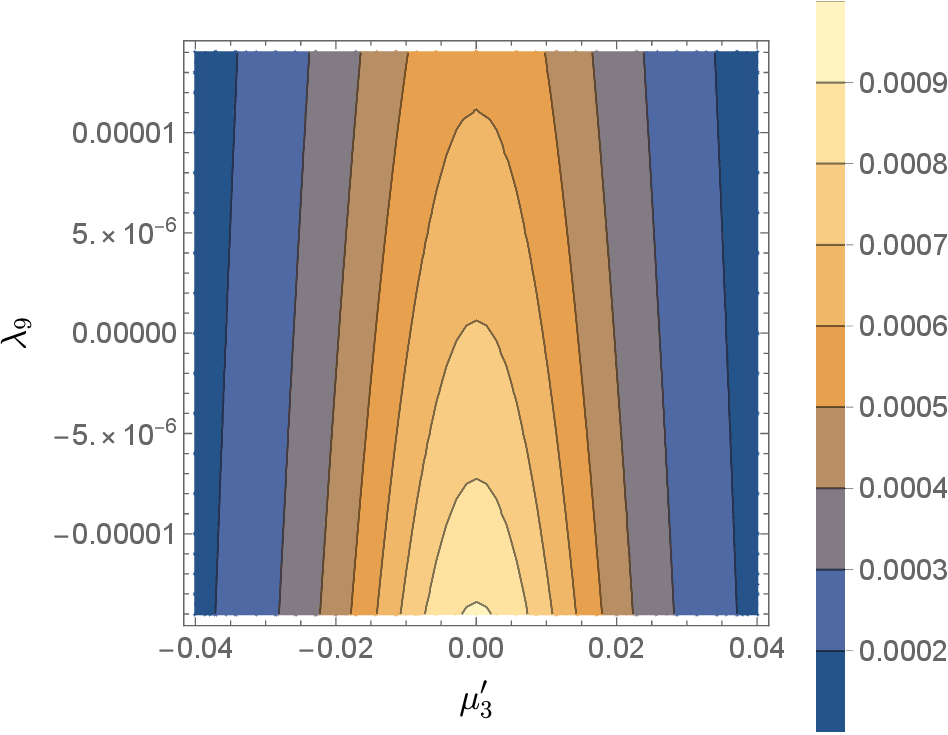}
    \end{tabular}
    \caption{Contour plots show the configuration of $\rho$ in four planes. The top panel is shown in the $\mu_{67} - \l_7$ plane for $\l_7>0$ (left) and $\l_7<0$ (right). The middle panel is shown in the $\mu_{67} - \l_9$ plane with $\l_9>0$ (left) and $\l_9<0$ (right). Meanwhile, the bottom panel shows the configuration in the $\mu'_3 - \l_7$ and $\mu'_3 - \l_9$ planes (from left to right) for $\l_9>0$ and $\l_7>0$, respectively. We take a set values of the fixed parameter as $\kappa = 4$, $\mu'_3 = (-10^{-2},10^{-2})$, $\mu_{67} = -10^{-6}$, $\l_7 = (-10^{-3},10^{-3})$, and $\l_9 = (-10^{-4},10^{-4}-10^{-3})$. In the bottom panel, the negative region of the parameter $\rho$ is not shown.}
    \label{fig:12}
\end{figure}

In this section, we numerically study the positivity conditions for the corresponding VEVs of the three scalar fields. The VEVs depend on parameters such as bare-mass coupling, cubic interaction coupling, and four-vertex interaction couplings. After scanning the parameter space of the model, we investigate the obtained mass dependence on the parameters of the model.

To investigate the positivity conditions of the VEVs obtained in Eq.\eqref{eq:19}-\eqref{eq.42}, it is convenient to introduce the following dimensionless parameters scaled by the VEV of the scalar $H$,
\begin{eqnarray}
    \kappa = \frac{v_1}{v_2},\label{kappaori} \\
    \rho = \frac{v_3}{v_2}. \label{rhoori}
\end{eqnarray}
By using the above notations, Eq.\eqref{eq:19}-\eqref{eq.42} alter into
\begin{align}
    \kappa &= \sqrt{\frac{\left(\mu'^2_5-\mu_{78}\right)}{\l_6}}; \label{kap12}  
    \sqrt{\frac{2\left(\mu'^2_2-\mu_{64}\right)}{\l_8}}, \\
    \rho &= \frac{-\mu_{67}}{2\sqrt{2} \mu'^2_3 + \l_{79}}. \label{rho}
\end{align}
where we have used the following simplified parameters as follows,
\begin{align}
    \mu_{78} &= \frac{1}{2} \left(\mu'_{7} \rho \sqrt{2} + \l_8 \right), \\
    \mu_{64} &= \frac{\mu'_6}{\sqrt{2}}\rho + \l_4, \\
    \mu_{67} &= \mu'_6 + \mu'_7 \kappa^2, \\
    \l_{79} &= \sqrt{2} \left( \l_7 + \l_9 \kappa^2 \right),\label{lam79} \\
    \mu_{i}' &= \frac{\mu_{i}}{v_2}\quad (i=2,3,5,6,7).
\end{align}
In the above equations, we define dimensionless coupling $\mu_{i}'$ to denote the bare-mass coupling $\mu_i$ scaled by the $v_2$, while the other couplings remain the same. In the numerical simulation, we do not specify the unit of parameters. Note that the numerical values for the dimensionless quantities, such as the ratio of VEV or ratio coupling over VEV, do not depend on the choice of the unit as long as the quantities in the ratio are given in the same unit. Thus in the simulation, we assume the hierarchy of $\rho\ll \kappa$ to be satisfied, following the hierarchy of the VEVs stated in Section \ref{sec2}, $v_1 > v_2 \gg v_3$, after using relations in Eqs.\eqref{kappaori} and \eqref{rhoori}. We also choose the value of $v_2$ approximately to be the VEV of the SM Higgs, $v_2\simeq 246$ GeV. 
 
Figure \ref{fig:2} shows the parameter $\kappa$ as a function of $\mu'_5$ where we have used the first function $\kappa$ in Eq.\eqref{kap12}. While we investigate the dependence of parameter $\mu'_5$, we set three possible values for $\mu_{78}$ and two different values of $\lambda_{6}$. We find that four possible conditions lead to the positive value of $\kappa$. One interesting finding is the case in $0<\kappa <1$ area with varied features. We set several values of the fixed parameters as $\mu_{78}=(-10^{-2},0,10^{-2})$ and $\lambda_{6}=(-10^{-2},10^{-2})$. In the case with $\mu_{78}=-10^{-2}$ and  $\l_6=10^{-2}$, $\kappa$ is always positive (solid red curve). Another interesting finding is when we take $\mu_{78}=10^{-2}$ but $\l_6=-10^{-2}$, either zero or positive of $\kappa$ are observed within $\mu'_5\in(-0.1,0.1)$. 

In Fig. \ref{fig:9}, we show the configurations of $\kappa$ for various values of the parameters $\mu_{78}$, $\l_6$, and $\mu'_5$, from top to bottom panels sequentially. Note that in the bottom panel, we only show the case with $\mu'_5>0$ since the negative value one will give the same plot ($\mu'_5$ has a quadratic form as appears in Eq.\eqref{kap12}). The line between the two-colored areas of the contour plots shows the varied value of $\kappa$. The white area is the non-allowed region and leads to the vanishing value of $\kappa$, whereas the colored area is the allowed region for the positive values.

Similarly, in Figure \ref{fig:3}, we show the parameter $\kappa$ dependence on $\mu'_{2}$ for the second function of Eq.\eqref{kap12}. We also find four possible conditions that lead
to the positive value of $\kappa$. In general, this figure has a similar feature as in Fig. \ref{fig:2}. The distinctive notable finding is the case with range $0<\kappa<5$ where in these bounds, the parameter $\kappa$ has more varied as shown by dotted blue, dot-dashed black, and dashed magenta lines from the figure even if $\kappa>1$ or $v_1 > v_2$. Figure \ref{fig:10} shows the configuration of $\kappa$ for various values of parameters $\mu'_{2}$, $\mu_{64}$, and $\lambda_{8}$ in three different types of contour plot planes. It has a similar property as in Fig. \ref{fig:9}. Namely, we show the non-allowed region (white area) leading to the vanishing of the parameter $\kappa$, while the allowed region is given in the colored area. We also observe some possible values of the parameters $\mu_{64}$ and $\lambda_{8}$ that can be chosen when determining the mass scale of the second scalar.

The parameter $\rho$ as a function of dimensionless coupling $\mu'_{3}$ is shown in Figure \ref{fig:4}. From this figure, we found two possible conditions leading to the positive value of the parameter $\rho$. There was an interesting plot when $\mu_{67},\lambda_{79}<0$, namely, within the small range of $\mu'_{3}$, it leads to the negative value of $\rho$, while other than that range leading to a positive one, as shown in the dotted-dashed magenta line. This could happen and can be easily checked from Eq. \eqref{rho}, when the value of $\lambda_{9}$ approaches $2\sqrt{2}{\mu'_{3}}^{2}$. In this case, $\lambda_{79}$ is mainly proportional to $\lambda_{9}$ since the term with $\lambda_{9}$ is multiplied with a factor $\kappa^{2}$ (see Eq.\eqref{lam79}). We also show contour plots for the configuration of the parameter $\rho$ in Figs. \ref{fig:11}-\ref{fig:12}. From these figures, we found that the darker-colored areas lead to negative values of $\rho$, while the brighter-colored areas provide positive ones. 

\begin{figure}[!tb]
    \centering
    \begin{tabular}{cc}
     \includegraphics[width=8cm]{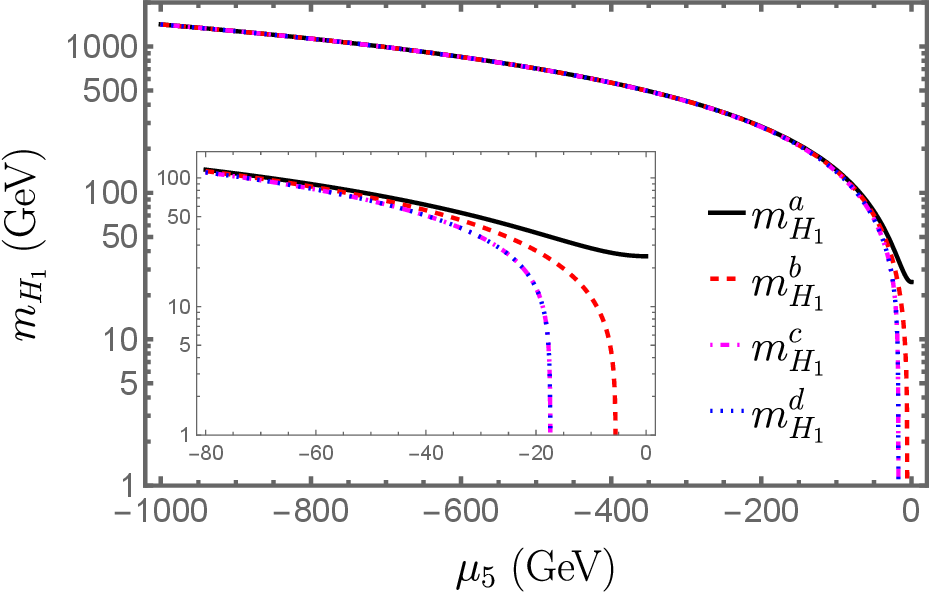} & \includegraphics[width=8cm]{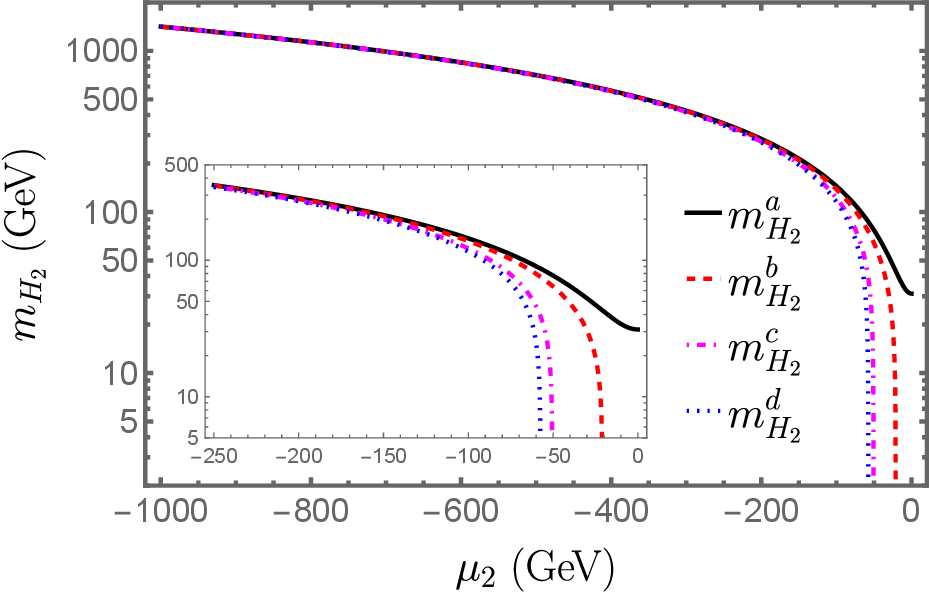} \\
     \multicolumn{2}{c}{\includegraphics[width=8cm]{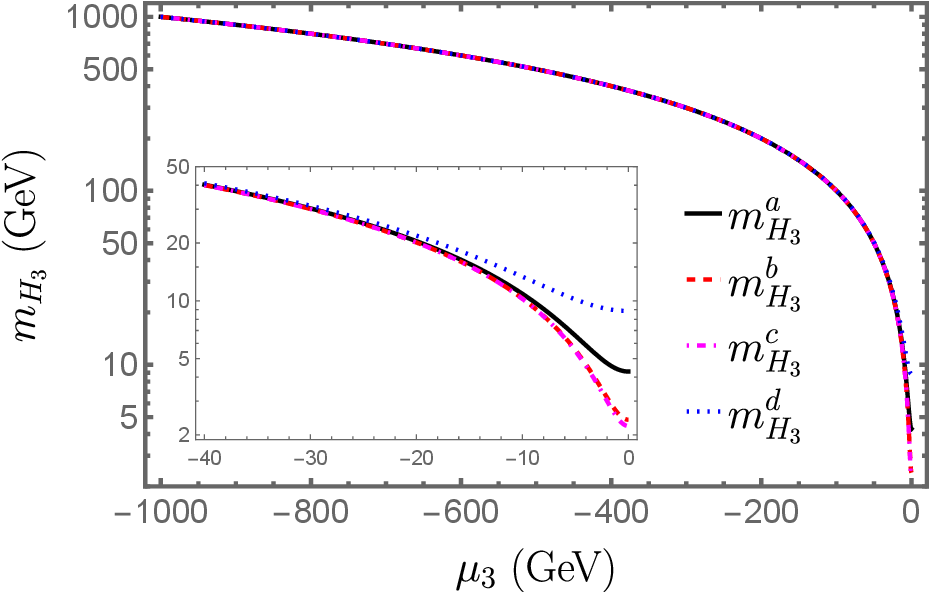}}
    \end{tabular}
    \caption{Dependence of $m_{H_i}$ on their full-dimensional couplings $\mu_i$, for four sets of the benchmark points. The upper left, upper right, and lower panel show the dependence of $m_{H_1}$ versus $\mu_5$, $m_{H_2}$ versus $\mu_2$, and $m_{H_3}$ versus $\mu_3$, respectively. The solid black, dashed red, dot-dashed magenta, and dotted blue lines, respectively correspond to the three Higgs masses at the chosen benchmark $a$, $b$, $c$, and $d$ given in Eqs. \eqref{eq46}-\eqref{eq49}.}
    \label{fig:5}
\end{figure}

\begin{figure}[!tb]
    \centering
    \begin{tabular}{cc}
        \includegraphics[width=7cm]{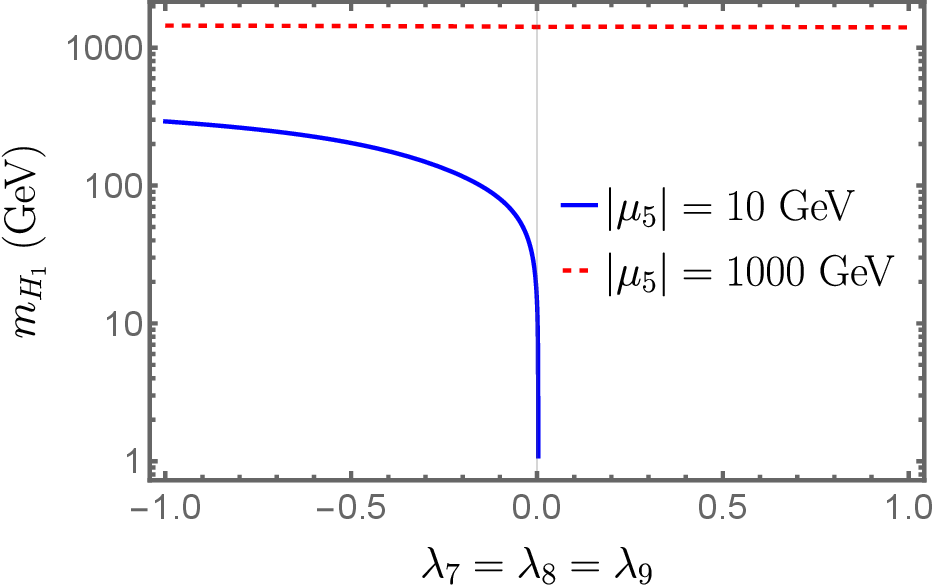} & \includegraphics[width=7cm]{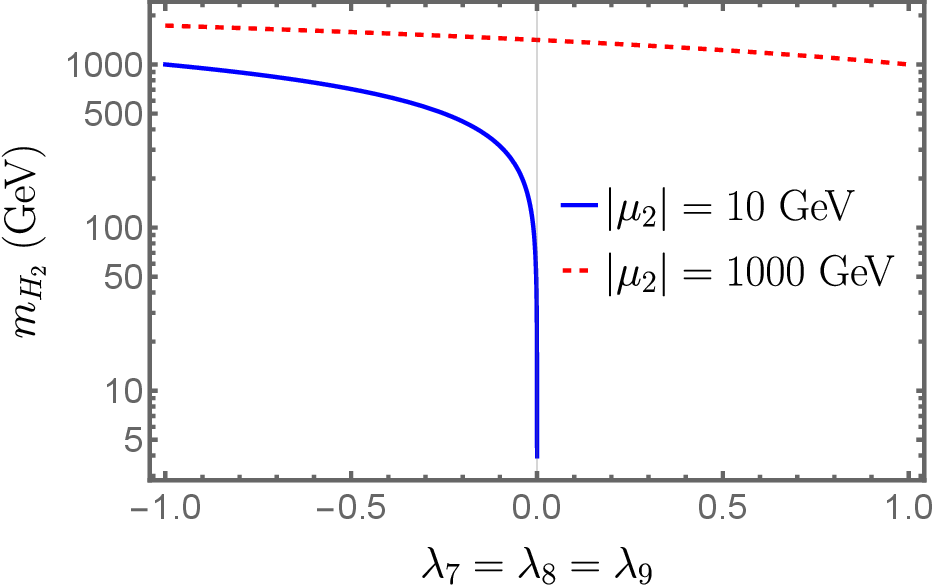} \\
        \multicolumn{2}{c}{\includegraphics[width=7cm]{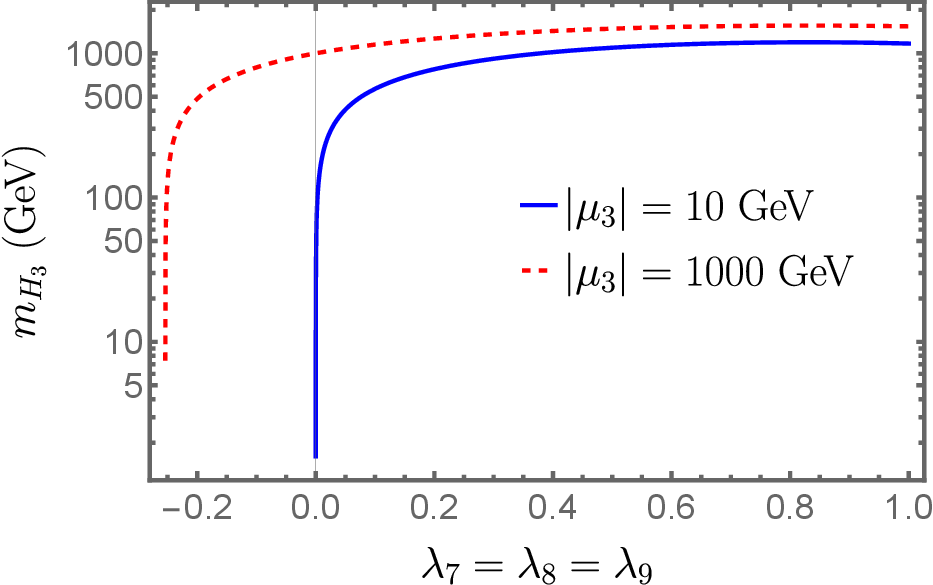}}
    \end{tabular}
    \caption{Dependence of $m_{H_i}$ on coupling $\l_7=\l_8=\l_9$. We set $\l_4=\l_6=0.129$. The solid blue and dashed red lines correspond to the functions of $m_{H_i}$ taking $|\mu_i|=10$ GeV and $|\mu_i|=1000$ GeV ($i=2,3,5$), respectively.}
    \label{fig:13}
\end{figure}

Next, we investigate the obtained masses of the scalar Higgs after the scanning parameter space dependent on the potential. Since we have assumed that the second scalar is the SM-like Higgs with VEV $v_{2}\simeq 246$ GeV, we perform the following simulations with a fixed $\lambda_{4}=0.129$. We choose the value of parameters, $\kappa\simeq4$ and $\rho \simeq 1\times 10^{-4}$, correspondingly $v_1\sim$ 1 TeV and $v_3\sim$ 10$^{-2}$  GeV, and take four different benchmark points as follows,  
\begin{align}
    &\text{benchmark }a\text{:}
    \begin{cases}
        & (\mu_2, \mu_3, \mu_5) = (0.220,0.020,4.0)v_2, \\
        & (\mu_6, \mu_7) = (-10^{-7},-10^{-7}),\\
        & (\l_6, \l_7, \l_8, \l_9) = (1.00, -10^{-3}, -10^{-2}, 10^{-4}).\\
    \end{cases} \label{eq46}\\
    &\text{benchmark }b\text{:}
    \begin{cases}
        & (\mu_2, \mu_3, \mu_5) = (0.370,0.010,3.8)v_2, \\
        & (\mu_6, \mu_7) = (-10^{-6},-10^{-6}),\\
        & (\l_6, \l_7, \l_8, \l_9) = (0.80, 10^{-5}, 10^{-3}, 10^{-5}).\\
    \end{cases} \label{eq47}\\
    &\text{benchmark }c\text{:}
    \begin{cases}
        & (\mu_2, \mu_3, \mu_5) = (0.415,0.020,2.9)v_2, \\
        & (\mu_6, \mu_7) = (-10^{-6},-10^{-6}),\\
        & (\l_6, \l_7, \l_8, \l_9) = (0.54, 10^{-5}, 10^{-2}, 10^{-5}).\\
    \end{cases} \label{eq48} \\
    &\text{benchmark }d\text{:}
    \begin{cases}
        & (\mu_2, \mu_3, \mu_5) = (0.430,0.001,3.3)v_2, \\
        & (\mu_6, \mu_7) = (-10^{-5},-10^{-5}),\\
        & (\l_6, \l_7, \l_8, \l_9) = (0.60, 10^{-3}, 10^{-2}, 10^{-4}),\\
    \end{cases} \label{eq49}
\end{align}
so that they satisfy the positivity conditions. 
In the above expressions, the parameters $\mu_2,\mu_{3}$ and $\mu_{5}$ are scaled with the VEV of the second scalar Higgs. 

In Figure \ref{fig:5}, we show the obtained mass dependence $m_{H_{i}}$ (see Eqs. \eqref{eq.8}-\eqref{eq.8c}) on their coupling parameters $\mu_{i}$ using benchmark points in Eqs. \eqref{eq47}-\eqref{eq49}. We note that there is a strong dependence of the $\lambda_7$ and $\l_{8}$ when parameters $\mu_5$, $\mu_2$, and $\mu_3$ decrease. Namely, either they are positive or negative values, in the region $|\mu_{5}|<80$ GeV (top-left panel), $|\mu_{2}|<250$ GeV (top-right panel), and $|\mu_{3}|<40$ GeV (bottom panel), they would have very different behavior as shown in the solid black line of Fig. \ref{fig:5}. Figure \ref{fig:13} shows the dependence of mass $m_{H_{i}}$ as a function of coupling $\lambda_{7}=\l_{8}=\l_{9}$ for two different values of $|\mu_{i}|$ ($i=2,3,5$). We have assumed that all of the mixing interactions among the scalars have the same strength. We also set the coupling strength of the SM-like scalar as strong as the coupling of the first heavy scalar, $\l_4=\l_6=0.129$.

Below we discuss the recent experimental bounds in connection with the result of our numerical study. From the chosen benchmark points in Eqs.\eqref{eq47}-\eqref{eq49}, we summarize the mass range of the three scalars Higgs in Table \ref{table:2}. The current searches at LHC give some constraints for the masses of the massive scalars in various decay channels. The search for a massive scalar in the resonant process $X\rightarrow Y H \rightarrow b\bar{b}b\Bar{b}$ is done by CMS collaboration with the scalar Higgs mass region, $0.9 < m_{X}<4$ TeV and $60< m_{Y}<600$ GeV at 95\% confidence level and in the range from 0.1 fb to 150 fb \cite{CMS:2022suh}. Another search has been also performed by ATLAS collaboration for a  resonant and non-resonant Higgs boson pair production in the $b\bar{b}\tau^{+}\tau^{-}$ decay channel. A broad excess is observed in the both channels, $\tau_{\text{had}}\tau_{\text{had}}$ and $\tau_{\text{lep}}\tau_{\text{had}}$, with the mass range between $700$ GeV and $1.2$ TeV \cite{ATLAS:2022xzm}. The most significant combined excess is at a signal mass hypothesis of 1 TeV with a local significance of $3.1\sigma$ and a global significance of $2.0\sigma$. Previous searches of different channels were performed by ATLAS and CMS for high \cite{ATLAS:2015pre,CMS:2018amk,ATLAS:2019erb,ATLAS:2019qdc,ATLAS:2020zms}, and low mass \cite{CMS:2019spf,CMS:2020ffa,CMS:2022suh,ATLAS:2018emt,ATLAS:2021hbr,ATLAS:2021ldb} region of Higgs boson.
Refs. \cite{Robens:2019kga,Robens:2022nnw} studied the phenomenology of a model with the spirit of two real-scalar singlet extensions. In this particular model, the masses of the scalars and their VEVs are set to be within 1 GeV $\leq m_{h_1}, m_{h_2}, v_{h_1}, v_{h_2} \leq$ 1 TeV. We note that the fifth benchmark (BP5) case of their work is comparable to our study, in which their heavy $h_1$ scalar is analog to the $h_1$ scalar in our study whereas their $h_2$ scalar is analog to our light $h_3$ scalar. This scenario is reported to be constrained by a recent ATLAS search for $pp\rightarrow h_1\rightarrow h_2 h_2\rightarrow 4\gamma$ signature with the mass range, $m_{h_1} \geq 200$ GeV and $0.1 \text{ GeV} \leq m_{h_2} \leq 10 \text{ GeV}$ \cite{ATLAS:2018dfo}.

\begin{table}
    \centering
    \begin{tabular}{ccccc}
        \hline \hline
        \text{Benchmark} & $m_{H_1}$ (TeV) & $m_{H_2}$ (GeV) & $m_{H_3}$ (GeV) & $M_{Z'}$ (TeV)\\
        \hline
        $a$ & 1.39 & 124.7 & 6.52 & 0.98  \\
        $b$ & 1.32 & 125.3 & 3.44 & 1.05 \\
        $c$ & 1.01 & 125.5 & 5.40 & 0.97 \\
        $d$ & 1.15 & 125.7 & 8.85 & 1.05 \\
        \hline \hline
    \end{tabular}
    \caption{The masses of the three scalars and the new neutral boson $Z'$ with fixed $g''\simeq 1$ using the benchmarks in Eqs. \eqref{eq46}-\eqref{eq49}.}
    \label{table:2}
\end{table}

\begin{figure}[ht]
    \centering
    \includegraphics[width=8.5cm]{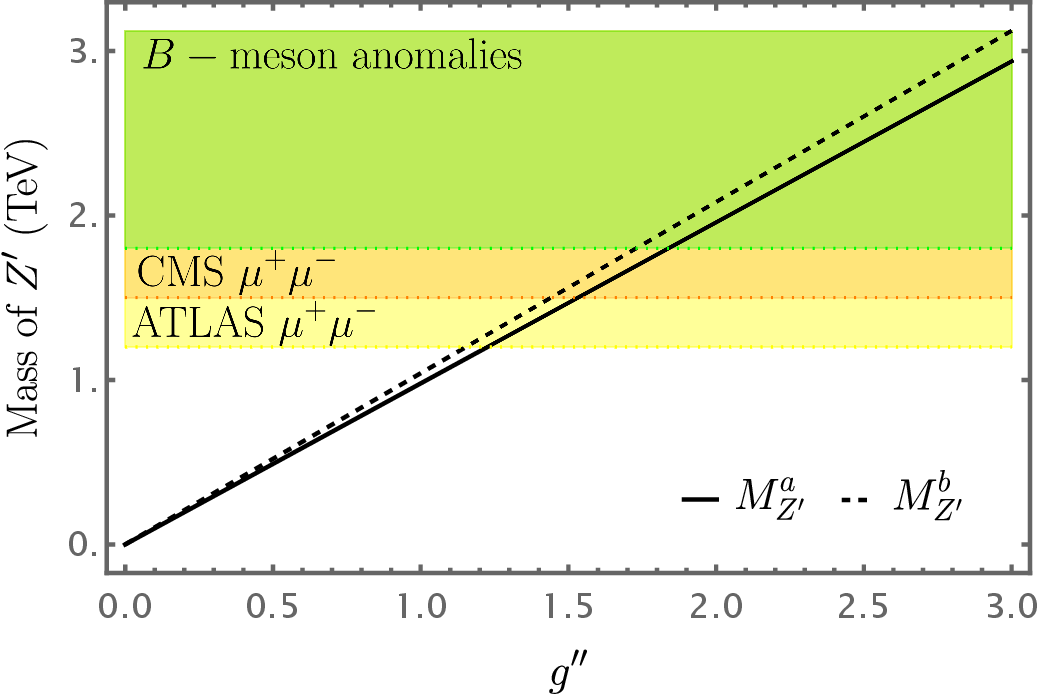}
    \caption{Dependence of the $Z'$ mass on its coupling $g''$. The solid black and dashed black lines correspond to the $M_{Z'}$ at the chosen benchmark $a$ and $b$ given in Eqs. \eqref{eq46} and \eqref{eq47}, respectively. The yellow and orange colored regions subsequently show the mass range of $M_{Z'}$ excluded by ATLAS ($>1.2$ TeV) \cite{ATLAS:2019erb} and CMS ($>1.5$ TeV) \cite{CMS:2021ctt} for di-muon experimental searches within $U(1)_{X}$ extension models \cite{Allanach:2019mfl,Allanach:2021bbd}, while the green region ($>1.8$ TeV) shows the disfavoured $Z'$ mass in the $B$-meson anomalies by taking the constraints from $B_s - \Bar{B}_s$ mixing and imposing the perturbative requirements \cite{Xu:2018pnq,Bause:2021prv}.}
    \label{fig:14}
\end{figure}

In Figure \ref{fig:14}, we show $Z'$ mass dependence on its coupling $g''$ as in Eq.\eqref{gaugeboson} and it behaves linearly with respect to $g''$ for fixed $v_1$. We only use two benchmarks, $a$ and $b$ as shown in Eqs. \eqref{eq46} and \eqref{eq47}. In the figure, the yellow ($M_{Z'}>1.2$ TeV) and the orange ($M_{Z'}>1.5$ TeV) regions are excluded at 95\% confidence level by the ATLAS \cite{ATLAS:2019erb} and CMS \cite{CMS:2021ctt} di-muon search, in which the parameter space is constrained to fit the $b\rightarrow s\mu^{+}\mu^{-}$ anomalies in the model with $U(1)_X$ extensions \cite{Allanach:2019mfl,Allanach:2021bbd}. The green region shows the mass range of $M_{Z'}$ excluded at $M_{Z'}>1.8$ TeV \cite{Xu:2018pnq,Bause:2021prv} for $B$-meson anomalies fit by considering the constraints from $B_s - \Bar{B}_s$ mixing and imposing perturbative requirements on the Yukawa and gauge couplings. With the chosen benchmarks in Eqs.\eqref{eq47}-\eqref{eq49}, we find numerically that the $Z'$ mass is of order TeV scale which is still within the collider bounds \cite{Nadolsky:2008zw,Rizzo:2014xma,Melnikov:2006kv,Montano-Dominguez:2022ytt,McEntaggart:2022hey,Gutierrez-Rodriguez:2015qka,Gutierrez-Rodriguez:2018jfl,Ramirez-Sanhez:2018xnh}, taking the fixed coupling $g''$ approximately to be unity (Table \ref{table:2}).
Regarding the gauge coupling of the $Z'$, in \cite{Allanach:2018lvl}, they have predicted that the values of the coupling reach 8.4 when the total width of $Z'$ is equal to its mass for the Third Family Hypercharge Model. Beyond this value, the model enters a non-perturbative regime, namely, the total width is larger than the $Z'$ mass. In other studies, they pointed out that the best-fit point of the value of $Z'$ coupling, $g_{Z'}=0.418\times (M_{Z'}/3\ \text{TeV})$ for $Y_3$ model, where $g_{Z'}$ has been scaled linearly \cite{Allanach:2021kzj}. The coupling value can be also much more smaller, $g_{Z'}=0.15$ for $M_{Z'}=3$ TeV, for di-muon resonance in the $U(1)_{B_{3}-L_{2}}$ extension model as has been explored in \cite{Allanach:2020kss}. The detailed investigation of the collider phenomenology of the $Z'$ in this proposed model lies beyond the scope of this study.
\section{Summary and Outlook} \label{sec5}
In this work, we have investigated scalar and gauge boson sectors in an extended SM with additional $U(1)_D$ symmetry. The masses of both scalar and gauge bosons are obtained through a spontaneous symmetry breaking of the scalar fields that have non-zero VEV. Their masses are written in terms of the coupling and parameters of the potential. In addition to the usual electroweak neutral gauge boson, we have obtained a new gauge boson $Z'$ in which the mass scale will be determined by its corresponding coupling and the VEV scale of the scalar Higgs $\Phi'$. 

We have also numerically studied the positivity conditions for the VEV of the obtained scalars. We have introduced dimensionless parameters in which all of the dimension-full parameters of the potential are scaled by the VEV of the second scalar. Thus the relations between the VEVs and the other parameters of the potential are more straightforward. It enables us to investigate the scanning of the parameter space in various settings. We then take several benchmark points within the allowed region of the parameter space and perform simulations for the obtained masses concerning their corresponding bare-mass couplings.

As an outlook of this work, the obtained new gauge boson particle could be a signature for a new physics candidate and may be tested in the near future experiment. A detailed analysis of the vacuum structure for this model is also needed. This analysis could help us to obtain the constraint on the model parameters from collider data on the obtained mass of the Higgs scalars. This work will be done in the future.

\vspace{1cm}

\noindent {\bf Acknowledgment}

We would like to thank the theoretical high energy physics research group of the Research Center for Quantum Physics for their kind hospitality. Y.K.A. thanks the particle physics research internship program of Merdeka Belajar Kampus Merdeka (MBKM) at BRIN where this work was initiated. This work was supported by the State University of Malang under Grant No. 5.4.1/UN32/KP/2023.
\appendix

\section{Appendix A} \label{app.A}
The diagonalization of the squared mass matrix obtained in Eq.\eqref{eq.6} is carried out in two stages \cite{Dutta:2022knf}.
The matrix is considered approximately as a block-diagonalized to obtain the exact diagonal mass matrix. For simplicity, we write symbolically the mass matrix in Eq.\eqref{eq.6} as follows,
\begin{equation}
    M^2_{h} = \begin{pmatrix}
        A & B & C \\
        B & D & E \\
        C & E & F \\
    \end{pmatrix}
\end{equation}
Applying a unitary matrix $M$ to the above matrix results in, 
\begin{equation}
    \begin{aligned}
        M^{\d} M^2_{h} M &=
        \begin{pmatrix}
            \c{\al} & 0 & - \s{\al} \\
            0 & \c{\al} & - \s{\al} \\
            \s{\al} & \s{\al} & \c{\al} \\
        \end{pmatrix}
        \begin{pmatrix}
            A & B & C \\
            B & D & E \\
            C & E & F \\
        \end{pmatrix}
        \begin{pmatrix}
            \c{\al} & 0 & \s{\al} \\
            0 & \c{\al} & \s{\al} \\
            - \s{\al} & - \s{\al} & \c{\al} \\
        \end{pmatrix} \\
        &= \begin{pmatrix}
            A' & B' & C' \\
            B' & D' & E' \\
            C' & E' & F' \\
        \end{pmatrix}
    \end{aligned}
\end{equation}
where
\begin{equation}
        \begin{aligned}
        A' &= A \c^2{\al} + F \s^2{\al} - C \s{2\al} \\
        B' &= B \c^2{\al} + F \s^2{\al} - \frac{1}{2}(C+E) \s{2\al} \\
        C' &= C \c^2{\al} + (A+B) \s{\al} \c{\al} - F \s{\al} \c{\al} - (C+E) \s^2{\al} \\
        D' &= D \c^2{\al} + F \s^2{\al} - E \s{2\al} \\
        E' &= E \c^2{\al} + (B+D) \s{\al} \c{\al} - F \s{\al} \c{\al} - (C+E) \s^2{\al} \\
        F' &= F \c^2{\al} + (A+2B+D) \s^2{\al} + (C+E) \s{2\al}.
    \end{aligned}
\end{equation}
Due to rotation by the matrix $M$, the values of the angle $\al$ [ignoring $\pazocal{O}(\sin^2{\al})$] are obtained as,    
\begin{align} \label{eq.15}
    \tan{\al} = \frac{C}{F-A-B} 
    && \mathrm{or} &&
    \tan{\al} = \frac{E}{F-B-D}
\end{align}
which leads to the value of $\tan{\al}$ given in Eq.\eqref{eq.tanAl}. Now the block-diagonalized mass matrix can then be rewritten as,
\begin{equation} \label{eq.16}
    M'^{2}_{h} = \begin{pmatrix}
        A' & B' & 0 \\
        B' & D' & 0 \\
        0 & 0 & F' \\
    \end{pmatrix}.
\end{equation}

In the second stage, the exact diagonalization is done using the Euler rotation matrix $R$ in Eq.\eqref{eq.7}. Thus, we have
\begin{equation}
    \begin{aligned}
        R^{\d} M'^{2}_{h} R &=
        \begin{pmatrix}
            \c{\beta} & \s{\beta} & 0 \\
            - \s{\beta} & \c{\beta} & 0 \\
            0 & 0 & 1 \\
        \end{pmatrix}
        \begin{pmatrix}
            A' & B' & 0 \\
            B' & D' & 0 \\
            0 & 0 & F' \\
        \end{pmatrix}
        \begin{pmatrix}
            \c{\beta} & - \s{\beta} & 0 \\
            \s{\beta} & \c{\beta} & 0 \\
            0 & 0 & 1 \\
        \end{pmatrix} \\
        &= \begin{pmatrix}
            A'' & B'' & 0 \\
            B'' & D'' & 0 \\
            0 & 0 & F'' \\
        \end{pmatrix},
    \end{aligned}
\end{equation}
where
\begin{equation}
    \begin{aligned}
        A'' &= A' \c^2{\beta} + B' \s{2\beta} + D' \s^2{\beta}, \\
        B'' &= B' \c^2{\beta} + \frac{1}{2}(D'-A') \s{2\beta} - B' \s^2{\beta},\\
        D'' &= D' \c^2{\beta} - B' \s{2\beta} + A' \s^2{\beta}, \\
        F'' &= F'.
    \end{aligned}
\end{equation}
Due to rotation by the matrix $R$, then the $h_1$-$h_2$ mixing angle, $\beta$, can also be obtained as,
\begin{equation} \label{eq.17}
    \tan{2\beta} = \frac{2B'}{A'-D'} ,
\end{equation}
and it leads to the mixing angle given in Eq.\eqref{eq.tanB}. The diagonalized squared mass matrix then can be written as,
\begin{equation} \label{eq:26}
    M''^2_h = m^2_{H_i} =
    \begin{pmatrix}
            m^2_{H_1} & 0 & 0 \\
            0 & m^2_{H_2} & 0 \\
            0 & 0 & m^2_{H_3} \\
    \end{pmatrix},
\end{equation}   
with
\begin{equation}
    \begin{aligned}
        m^2_{H_1} &= A' \c^2{\beta} + D' \s^2{\beta} + B' \s{2\beta}, \\
        m^2_{H_2} &= D' \c^2{\beta} + A' \s^2{\beta} - B' \s{2\beta}, \\
        m^2_{H_3} &= F \c^2{\al} + (A+2B+D) \s^2{\al} + (C+E) \s{2\al}. \\
    \end{aligned}
\end{equation}
Substituting the explicit symbolical expression in the above equation leads to the mass of the scalar Higgs in the diagonal form given in Eqs.\eqref{eq.8}-\eqref{eq.8c}.



\begin{thebibliography}{00}
\bibitem{ATLAS:2012yve}
{\bfseries ATLAS} Collaboration, G.~Aad {\em et~al.}, ``{Observation of a new
	particle in the search for the Standard Model Higgs boson with the ATLAS
	detector at the LHC},''
\href{https://dx.doi.org/10.1016/j.physletb.2012.08.020}{{\em Phys. Lett. B}
	{\bfseries 716} (2012) 1--29},
\href{https://arxiv.org/abs/1207.7214}{{\ttfamily arXiv:1207.7214 [hep-ex]}}.

\bibitem{CMS:2012qbp}
{\bfseries CMS} Collaboration, S.~Chatrchyan {\em et~al.}, ``{Observation of a
	New Boson at a Mass of 125 GeV with the CMS Experiment at the LHC},''
\href{https://dx.doi.org/10.1016/j.physletb.2012.08.021}{{\em Phys. Lett. B}
	{\bfseries 716} (2012) 30--61},
\href{https://arxiv.org/abs/1207.7235}{{\ttfamily arXiv:1207.7235 [hep-ex]}}.

\bibitem{ParticleDataGroup:2022pth}
{\bfseries Particle Data Group} Collaboration, R.~L. Workman {\em et~al.},
``{Review of Particle Physics},''
\href{https://dx.doi.org/10.1093/ptep/ptac097}{{\em PTEP} {\bfseries 2022}
	(2022) 083C01}.

\bibitem{Higgs:1964ia}
P.~W. Higgs, ``{Broken symmetries, massless particles and gauge fields},''
\href{https://dx.doi.org/10.1016/0031-9163(64)91136-9}{{\em Phys. Lett.}
	{\bfseries 12} (1964) 132--133}.

\bibitem{Englert:1964et}
F.~Englert and R.~Brout, ``{Broken Symmetry and the Mass of Gauge Vector
	Mesons},'' \href{https://dx.doi.org/10.1103/PhysRevLett.13.321}{{\em Phys.
		Rev. Lett.} {\bfseries 13} (1964) 321--323}.

\bibitem{Guralnik:1964eu}
G.~S. Guralnik, C.~R. Hagen, and T.~W.~B. Kibble, ``{Global Conservation Laws
	and Massless Particles},''
\href{https://dx.doi.org/10.1103/PhysRevLett.13.585}{{\em Phys. Rev. Lett.}
	{\bfseries 13} (1964) 585--587}.

\bibitem{Higgs:1966ev}
P.~W. Higgs, ``{Spontaneous Symmetry Breakdown without Massless Bosons},''
\href{https://dx.doi.org/10.1103/PhysRev.145.1156}{{\em Phys. Rev.}
	{\bfseries 145} (1966) 1156--1163}.

\bibitem{Kibble:1967sv}
T.~W.~B. Kibble, ``{Symmetry breaking in nonAbelian gauge theories},''
\href{https://dx.doi.org/10.1103/PhysRev.155.1554}{{\em Phys. Rev.}
	{\bfseries 155} (1967) 1554--1561}.

\bibitem{OConnell:2006rsp}
D.~O'Connell, M.~J. Ramsey-Musolf, and M.~B. Wise, ``{Minimal Extension of the
	Standard Model Scalar Sector},''
\href{https://dx.doi.org/10.1103/PhysRevD.75.037701}{{\em Phys. Rev. D}
	{\bfseries 75} (2007) 037701},
\href{https://arxiv.org/abs/hep-ph/0611014}{{\ttfamily
		arXiv:hep-ph/0611014}}.

\bibitem{Bowen:2007ia}
M.~Bowen, Y.~Cui, and J.~D. Wells, ``{Narrow trans-TeV Higgs bosons and H
	---\ensuremath{>} hh decays: Two LHC search paths for a hidden sector Higgs
	boson},'' \href{https://dx.doi.org/10.1088/1126-6708/2007/03/036}{{\em JHEP}
	{\bfseries 03} (2007) 036},
\href{https://arxiv.org/abs/hep-ph/0701035}{{\ttfamily
		arXiv:hep-ph/0701035}}.

\bibitem{Espinosa:2011ax}
J.~R. Espinosa, T.~Konstandin, and F.~Riva, ``{Strong Electroweak Phase
	Transitions in the Standard Model with a Singlet},''
\href{https://dx.doi.org/10.1016/j.nuclphysb.2011.09.010}{{\em Nucl. Phys. B}
	{\bfseries 854} (2012) 592--630},
\href{https://arxiv.org/abs/1107.5441}{{\ttfamily arXiv:1107.5441 [hep-ph]}}.

\bibitem{Pruna:2013bma}
G.~M. Pruna and T.~Robens, ``{Higgs singlet extension parameter space in the
	light of the LHC discovery},''
\href{https://dx.doi.org/10.1103/PhysRevD.88.115012}{{\em Phys. Rev. D}
	{\bfseries 88} no.~11, (2013) 115012},
\href{https://arxiv.org/abs/1303.1150}{{\ttfamily arXiv:1303.1150 [hep-ph]}}.

\bibitem{Haber:1978jt}
H.~E. Haber, G.~L. Kane, and T.~Sterling, ``{The Fermion Mass Scale and
	Possible Effects of Higgs Bosons on Experimental Observables},''
\href{https://dx.doi.org/10.1016/0550-3213(79)90225-6}{{\em Nucl. Phys. B}
	{\bfseries 161} (1979) 493--532}.

\bibitem{Gunion:2002zf}
J.~F. Gunion and H.~E. Haber, ``{The CP conserving two Higgs doublet model: The
	Approach to the decoupling limit},''
\href{https://dx.doi.org/10.1103/PhysRevD.67.075019}{{\em Phys. Rev. D}
	{\bfseries 67} (2003) 075019},
\href{https://arxiv.org/abs/hep-ph/0207010}{{\ttfamily
		arXiv:hep-ph/0207010}}.

\bibitem{Branco:2011iw}
G.~C. Branco, P.~M. Ferreira, L.~Lavoura, M.~N. Rebelo, M.~Sher, and J.~P.
Silva, ``{Theory and phenomenology of two-Higgs-doublet models},''
\href{https://dx.doi.org/10.1016/j.physrep.2012.02.002}{{\em Phys. Rept.}
	{\bfseries 516} (2012) 1--102},
\href{https://arxiv.org/abs/1106.0034}{{\ttfamily arXiv:1106.0034 [hep-ph]}}.

\bibitem{Gunion:1984yn}
J.~F. Gunion and H.~E. Haber, ``{Higgs Bosons in Supersymmetric Models. 1.},''
\href{https://dx.doi.org/10.1016/0550-3213(86)90340-8}{{\em Nucl. Phys. B}
	{\bfseries 272} (1986) 1}. [Erratum: Nucl.Phys.B 402, 567--569 (1993)].

\bibitem{Gunion:1986nh}
J.~F. Gunion and H.~E. Haber, ``{Higgs Bosons in Supersymmetric Models. 2.
	Implications for Phenomenology},''
\href{https://dx.doi.org/10.1016/0550-3213(86)90050-7}{{\em Nucl. Phys. B}
	{\bfseries 278} (1986) 449}. [Erratum: Nucl.Phys.B 402, 569--569 (1993)].

\bibitem{Csaki:1996ks}
C.~Csaki, ``{The Minimal supersymmetric standard model (MSSM)},''
\href{https://dx.doi.org/10.1142/S021773239600062X}{{\em Mod. Phys. Lett. A}
	{\bfseries 11} (1996) 599},
\href{https://arxiv.org/abs/hep-ph/9606414}{{\ttfamily
		arXiv:hep-ph/9606414}}.

\bibitem{Heinemeyer:2004ms}
S.~Heinemeyer, ``{MSSM Higgs physics at higher orders},''
\href{https://dx.doi.org/10.1142/S0217751X06031028}{{\em Int. J. Mod. Phys.
		A} {\bfseries 21} (2006) 2659--2772},
\href{https://arxiv.org/abs/hep-ph/0407244}{{\ttfamily
		arXiv:hep-ph/0407244}}.

\bibitem{Draper:2016pys}
P.~Draper and H.~Rzehak, ``{A Review of Higgs Mass Calculations in
	Supersymmetric Models},''
\href{https://dx.doi.org/10.1016/j.physrep.2016.01.001}{{\em Phys. Rept.}
	{\bfseries 619} (2016) 1--24},
\href{https://arxiv.org/abs/1601.01890}{{\ttfamily arXiv:1601.01890
		[hep-ph]}}.

\bibitem{Lopez-Val:2013yba}
D.~L\'opez-Val, T.~Plehn, and M.~Rauch, ``{Measuring extended Higgs sectors as
	a consistent free couplings model},''
\href{https://dx.doi.org/10.1007/JHEP10(2013)134}{{\em JHEP} {\bfseries 10}
	(2013) 134}, \href{https://arxiv.org/abs/1308.1979}{{\ttfamily
		arXiv:1308.1979 [hep-ph]}}.

\bibitem{Chen:2014ask}
C.-Y. Chen, S.~Dawson, and I.~M. Lewis, ``{Exploring resonant di-Higgs boson
	production in the Higgs singlet model},''
\href{https://dx.doi.org/10.1103/PhysRevD.91.035015}{{\em Phys. Rev. D}
	{\bfseries 91} no.~3, (2015) 035015},
\href{https://arxiv.org/abs/1410.5488}{{\ttfamily arXiv:1410.5488 [hep-ph]}}.

\bibitem{Carena:2013ooa}
M.~Carena, I.~Low, N.~R. Shah, and C.~E.~M. Wagner, ``{Impersonating the
	Standard Model Higgs Boson: Alignment without Decoupling},''
\href{https://dx.doi.org/10.1007/JHEP04(2014)015}{{\em JHEP} {\bfseries 04}
	(2014) 015}, \href{https://arxiv.org/abs/1310.2248}{{\ttfamily
		arXiv:1310.2248 [hep-ph]}}.

\bibitem{BhupalDev:2014bir}
P.~S. Bhupal~Dev and A.~Pilaftsis, ``{Maximally Symmetric Two Higgs Doublet
	Model with Natural Standard Model Alignment},''
\href{https://dx.doi.org/10.1007/JHEP12(2014)024}{{\em JHEP} {\bfseries 12}
	(2014) 024}, \href{https://arxiv.org/abs/1408.3405}{{\ttfamily
		arXiv:1408.3405 [hep-ph]}}. [Erratum: JHEP 11, 147 (2015)].

\bibitem{Gu:2006dc}
P.-H. Gu and H.-J. He, ``{Neutrino Mass and Baryon Asymmetry from Dirac
	Seesaw},'' \href{https://dx.doi.org/10.1088/1475-7516/2006/12/010}{{\em JCAP}
	{\bfseries 12} (2006) 010},
\href{https://arxiv.org/abs/hep-ph/0610275}{{\ttfamily
		arXiv:hep-ph/0610275}}.

\bibitem{Barger:1987xw}
V.~D. Barger and K.~Whisnant, ``{Heavy $Z$ Boson Decays to Two Bosons in $E(6)$
	Superstring Models},''
\href{https://dx.doi.org/10.1103/PhysRevD.36.3429}{{\em Phys. Rev. D}
	{\bfseries 36} (1987) 3429}.

\bibitem{Espinosa:1997ji}
J.~R. Espinosa, ``{Z-prime gauge models from strings},''
\href{https://dx.doi.org/10.1016/S0920-5632(97)00656-7}{{\em Nucl. Phys. B
		Proc. Suppl.} {\bfseries 62} (1998) 187--196},
\href{https://arxiv.org/abs/hep-ph/9707541}{{\ttfamily
		arXiv:hep-ph/9707541}}.

\bibitem{Langacker:1991pg}
P.~Langacker and M.-x. Luo, ``{Constraints on additional $Z$ bosons},''
\href{https://dx.doi.org/10.1103/PhysRevD.45.278}{{\em Phys. Rev. D}
	{\bfseries 45} (1992) 278--292}.

\bibitem{Liao:2017uzy}
J.~Liao and D.~Marfatia, ``{COHERENT constraints on nonstandard neutrino
	interactions},''
\href{https://dx.doi.org/10.1016/j.physletb.2017.10.046}{{\em Phys. Lett. B}
	{\bfseries 775} (2017) 54--57},
\href{https://arxiv.org/abs/1708.04255}{{\ttfamily arXiv:1708.04255
		[hep-ph]}}.

\bibitem{Papoulias:2017qdn}
D.~K. Papoulias and T.~S. Kosmas, ``{COHERENT constraints to conventional and
	exotic neutrino physics},''
\href{https://dx.doi.org/10.1103/PhysRevD.97.033003}{{\em Phys. Rev. D}
	{\bfseries 97} no.~3, (2018) 033003},
\href{https://arxiv.org/abs/1711.09773}{{\ttfamily arXiv:1711.09773
		[hep-ph]}}.

\bibitem{Bobovnikov:2018fwt}
I.~D. Bobovnikov, P.~Osland, and A.~A. Pankov, ``{Improved constraints on the
	mixing and mass of $Z'$ bosons from resonant diboson searches at the LHC at
	$\sqrt{s}=13$ TeV and predictions for Run II},''
\href{https://dx.doi.org/10.1103/PhysRevD.98.095029}{{\em Phys. Rev. D}
	{\bfseries 98} no.~9, (2018) 095029},
\href{https://arxiv.org/abs/1809.08933}{{\ttfamily arXiv:1809.08933
		[hep-ph]}}.

\bibitem{Cheung:2022oji}
K.~Cheung and C.~J. Ouseph, ``{Constraining the Active-to-Heavy-Neutrino
	transitional magnetic moments associated with the $Z'$ interactions at
	FASER$\nu$},'' \href{https://arxiv.org/abs/2205.11077}{{\ttfamily
		arXiv:2205.11077 [hep-ph]}}.

\bibitem{Hewett:1988fe}
J.~L. Hewett and T.~G. Rizzo, ``{NEW NEUTRAL GAUGE BOSONS AT HIGH-ENERGY e+ e-
	COLLIDERS},'' \href{https://dx.doi.org/10.1142/S0217751X8900193X}{{\em Int.
		J. Mod. Phys. A} {\bfseries 4} (1989) 4551}.

\bibitem{ALEPH:2005ab}
{\bfseries ALEPH, DELPHI, L3, OPAL, SLD, LEP Electroweak Working Group, SLD
	Electroweak Group, SLD Heavy Flavour Group} Collaboration, S.~Schael {\em
	et~al.}, ``{Precision electroweak measurements on the $Z$ resonance},''
\href{https://dx.doi.org/10.1016/j.physrep.2005.12.006}{{\em Phys. Rept.}
	{\bfseries 427} (2006) 257--454},
\href{https://arxiv.org/abs/hep-ex/0509008}{{\ttfamily
		arXiv:hep-ex/0509008}}.

\bibitem{Ramirez-Sanchez:2016ugz}
F.~Ramirez-Sanchez, A.~Gutierrez-Rodriguez, and M.~A. Hernandez-Ruiz, ``{Higgs
	bosons production and decay at future $e^+e^-$ linear colliders as a probe of
	the B\textendash{}L model},''
\href{https://dx.doi.org/10.1088/0954-3899/43/9/095003}{{\em J. Phys. G}
	{\bfseries 43} no.~9, (2016) 095003},
\href{https://arxiv.org/abs/1606.04144}{{\ttfamily arXiv:1606.04144
		[hep-ph]}}.

\bibitem{Langacker:2008yv}
P.~Langacker, ``{The Physics of Heavy $Z^\prime$ Gauge Bosons},''
\href{https://dx.doi.org/10.1103/RevModPhys.81.1199}{{\em Rev. Mod. Phys.}
	{\bfseries 81} (2009) 1199--1228},
\href{https://arxiv.org/abs/0801.1345}{{\ttfamily arXiv:0801.1345 [hep-ph]}}.

\bibitem{Dutta:2022knf}
M.~Dutta, N.~Narendra, N.~Sahu, and S.~Shil, ``{Asymmetric self-interacting
	dark matter via Dirac leptogenesis},''
\href{https://dx.doi.org/10.1103/PhysRevD.106.095017}{{\em Phys. Rev. D}
	{\bfseries 106} no.~9, (2022) 095017},
\href{https://arxiv.org/abs/2202.04704}{{\ttfamily arXiv:2202.04704
		[hep-ph]}}.

\bibitem{Halzen:1984mc}
F.~Halzen and A.~D. Martin, {\em {Quarks And Leptons: An Introductory Course in
		Modern Particle Physics}}.
\newblock John Wiley \& Sons, Inc, {1984}.

\bibitem{CMS:2022suh}
{\bfseries CMS} Collaboration, A.~Tumasyan {\em et~al.}, ``{Search for a
	massive scalar resonance decaying to a light scalar and a Higgs boson in the
	four b quarks final state with boosted topology},''
\href{https://dx.doi.org/10.1016/j.physletb.2022.137392}{{\em Phys. Lett. B}
	{\bfseries 842} (2023) 137392},
\href{https://arxiv.org/abs/2204.12413}{{\ttfamily arXiv:2204.12413
		[hep-ex]}}.

\bibitem{ATLAS:2022xzm}
{\bfseries ATLAS} Collaboration, G.~Aad {\em et~al.}, ``{Search for resonant
	and non-resonant Higgs boson pair production in the $
	b\overline{b}{\tau}^{+}{\tau}^{-} $ decay channel using 13 TeV pp collision
	data from the ATLAS detector},''
\href{https://dx.doi.org/10.1007/JHEP07(2023)040}{{\em JHEP} {\bfseries 07}
	(2023) 040}, \href{https://arxiv.org/abs/2209.10910}{{\ttfamily
		arXiv:2209.10910 [hep-ex]}}.

\bibitem{ATLAS:2015pre}
{\bfseries ATLAS} Collaboration, G.~Aad {\em et~al.}, ``{Search for an
	additional, heavy Higgs boson in the $H\rightarrow ZZ$ decay channel at
	$\sqrt{s} = 8\;\text{ TeV }$ in $pp$ collision data with the ATLAS
	detector},'' \href{https://dx.doi.org/10.1140/epjc/s10052-015-3820-z}{{\em
		Eur. Phys. J. C} {\bfseries 76} no.~1, (2016) 45},
\href{https://arxiv.org/abs/1507.05930}{{\ttfamily arXiv:1507.05930
		[hep-ex]}}.

\bibitem{CMS:2018amk}
{\bfseries CMS} Collaboration, A.~M. Sirunyan {\em et~al.}, ``{Search for a new
	scalar resonance decaying to a pair of Z bosons in proton-proton collisions
	at $\sqrt{s}=13 $ TeV},''
\href{https://dx.doi.org/10.1007/JHEP06(2018)127}{{\em JHEP} {\bfseries 06}
	(2018) 127}, \href{https://arxiv.org/abs/1804.01939}{{\ttfamily
		arXiv:1804.01939 [hep-ex]}}. [Erratum: JHEP 03, 128 (2019)].

\bibitem{ATLAS:2019erb}
{\bfseries ATLAS} Collaboration, G.~Aad {\em et~al.}, ``{Search for high-mass
	dilepton resonances using 139 fb$^{-1}$ of $pp$ collision data collected at
	$\sqrt{s}=$13 TeV with the ATLAS detector},''
\href{https://dx.doi.org/10.1016/j.physletb.2019.07.016}{{\em Phys. Lett. B}
	{\bfseries 796} (2019) 68--87},
\href{https://arxiv.org/abs/1903.06248}{{\ttfamily arXiv:1903.06248
		[hep-ex]}}.

\bibitem{ATLAS:2019qdc}
{\bfseries ATLAS} Collaboration, G.~Aad {\em et~al.}, ``{Combination of
	searches for Higgs boson pairs in $pp$ collisions at $\sqrt{s} = $13 TeV with
	the ATLAS detector},''
\href{https://dx.doi.org/10.1016/j.physletb.2019.135103}{{\em Phys. Lett. B}
	{\bfseries 800} (2020) 135103},
\href{https://arxiv.org/abs/1906.02025}{{\ttfamily arXiv:1906.02025
		[hep-ex]}}.

\bibitem{ATLAS:2020zms}
{\bfseries ATLAS} Collaboration, G.~Aad {\em et~al.}, ``{Search for heavy Higgs
	bosons decaying into two tau leptons with the ATLAS detector using $pp$
	collisions at $\sqrt{s}=13$ TeV},''
\href{https://dx.doi.org/10.1103/PhysRevLett.125.051801}{{\em Phys. Rev.
		Lett.} {\bfseries 125} no.~5, (2020) 051801},
\href{https://arxiv.org/abs/2002.12223}{{\ttfamily arXiv:2002.12223
		[hep-ex]}}.

\bibitem{CMS:2019spf}
{\bfseries CMS} Collaboration, A.~M. Sirunyan {\em et~al.}, ``{Search for light
	pseudoscalar boson pairs produced from decays of the 125 GeV Higgs boson in
	final states with two muons and two nearby tracks in pp collisions at
	$\sqrt{s}=$ 13 TeV},''
\href{https://dx.doi.org/10.1016/j.physletb.2019.135087}{{\em Phys. Lett. B}
	{\bfseries 800} (2020) 135087},
\href{https://arxiv.org/abs/1907.07235}{{\ttfamily arXiv:1907.07235
		[hep-ex]}}.

\bibitem{CMS:2020ffa}
{\bfseries CMS} Collaboration, A.~M. Sirunyan {\em et~al.}, ``{Search for a
	light pseudoscalar Higgs boson in the boosted $\mu\mu\tau\tau$ final state in
	proton-proton collisions at $\sqrt{s}=$ 13 TeV},''
\href{https://dx.doi.org/10.1007/JHEP08(2020)139}{{\em JHEP} {\bfseries 08}
	(2020) 139}, \href{https://arxiv.org/abs/2005.08694}{{\ttfamily
		arXiv:2005.08694 [hep-ex]}}.

\bibitem{ATLAS:2018emt}
{\bfseries ATLAS} Collaboration, M.~Aaboud {\em et~al.}, ``{Search for Higgs
	boson decays into a pair of light bosons in the $bb\mu\mu$ final state in
	$pp$ collision at $\sqrt{s} = $13 TeV with the ATLAS detector},''
\href{https://dx.doi.org/10.1016/j.physletb.2018.10.073}{{\em Phys. Lett. B}
	{\bfseries 790} (2019) 1--21},
\href{https://arxiv.org/abs/1807.00539}{{\ttfamily arXiv:1807.00539
		[hep-ex]}}.

\bibitem{ATLAS:2021hbr}
{\bfseries ATLAS} Collaboration, G.~Aad {\em et~al.}, ``{Search for Higgs boson
	decays into a pair of pseudoscalar particles in the $bb\mu\mu$ final state
	with the ATLAS detector in $pp$ collisions at $\sqrt s$=13\,\,TeV},''
\href{https://dx.doi.org/10.1103/PhysRevD.105.012006}{{\em Phys. Rev. D}
	{\bfseries 105} no.~1, (2022) 012006},
\href{https://arxiv.org/abs/2110.00313}{{\ttfamily arXiv:2110.00313
		[hep-ex]}}.

\bibitem{ATLAS:2021ldb}
{\bfseries ATLAS} Collaboration, G.~Aad {\em et~al.}, ``{Search for Higgs
	bosons decaying into new spin-0 or spin-1 particles in four-lepton final
	states with the ATLAS detector with 139 fb$^{-1}$ of $pp$ collision data at
	$\sqrt{s}=13$ TeV},'' \href{https://dx.doi.org/10.1007/JHEP03(2022)041}{{\em
		JHEP} {\bfseries 03} (2022) 041},
\href{https://arxiv.org/abs/2110.13673}{{\ttfamily arXiv:2110.13673
		[hep-ex]}}.

\bibitem{Robens:2019kga}
T.~Robens, T.~Stefaniak, and J.~Wittbrodt, ``{Two-real-scalar-singlet extension
	of the SM: LHC phenomenology and benchmark scenarios},''
\href{https://dx.doi.org/10.1140/epjc/s10052-020-7655-x}{{\em Eur. Phys. J.
		C} {\bfseries 80} no.~2, (2020) 151},
\href{https://arxiv.org/abs/1908.08554}{{\ttfamily arXiv:1908.08554
		[hep-ph]}}.

\bibitem{Robens:2022nnw}
T.~Robens, ``{Two-Real-Singlet-Model Benchmark Planes},''
\href{https://dx.doi.org/10.3390/sym15010027}{{\em Symmetry} {\bfseries 15}
	no.~1, (2023) 27}, \href{https://arxiv.org/abs/2209.10996}{{\ttfamily
		arXiv:2209.10996 [hep-ph]}}.

\bibitem{ATLAS:2018dfo}
{\bfseries ATLAS} Collaboration, M.~Aaboud {\em et~al.}, ``{A search for pairs
	of highly collimated photon-jets in $pp$ collisions at $\sqrt{s}$ = 13 TeV
	with the ATLAS detector},''
\href{https://dx.doi.org/10.1103/PhysRevD.99.012008}{{\em Phys. Rev. D}
	{\bfseries 99} no.~1, (2019) 012008},
\href{https://arxiv.org/abs/1808.10515}{{\ttfamily arXiv:1808.10515
		[hep-ex]}}.

\bibitem{CMS:2021ctt}
{\bfseries CMS} Collaboration, A.~M. Sirunyan {\em et~al.}, ``{Search for
	resonant and nonresonant new phenomena in high-mass dilepton final states at
	$ \sqrt{s} $ = 13 TeV},''
\href{https://dx.doi.org/10.1007/JHEP07(2021)208}{{\em JHEP} {\bfseries 07}
	(2021) 208}, \href{https://arxiv.org/abs/2103.02708}{{\ttfamily
		arXiv:2103.02708 [hep-ex]}}.

\bibitem{Allanach:2019mfl}
B.~C. Allanach, J.~M. Butterworth, and T.~Corbett, ``{Collider constraints on
	$Z'$ models for neutral current B-anomalies},''
\href{https://dx.doi.org/10.1007/JHEP08(2019)106}{{\em JHEP} {\bfseries 08}
	(2019) 106}, \href{https://arxiv.org/abs/1904.10954}{{\ttfamily
		arXiv:1904.10954 [hep-ph]}}.

\bibitem{Allanach:2021bbd}
B.~C. Allanach and H.~Banks, ``{Hide and seek with the third family hypercharge
	model\textquoteright{}s $Z^\prime $ at the large hadron collider},''
\href{https://dx.doi.org/10.1140/epjc/s10052-022-10191-6}{{\em Eur. Phys. J.
		C} {\bfseries 82} no.~3, (2022) 279},
\href{https://arxiv.org/abs/2111.06691}{{\ttfamily arXiv:2111.06691
		[hep-ph]}}.

\bibitem{Xu:2018pnq}
F.-Z. Xu, W.~Zhang, J.~Li, and T.~Li, ``{Search for the vectorlike leptons in
	the U(1)$_X$ model inspired by the $B$-meson decay anomalies},''
\href{https://dx.doi.org/10.1103/PhysRevD.98.115033}{{\em Phys. Rev. D}
	{\bfseries 98} no.~11, (2018) 115033},
\href{https://arxiv.org/abs/1809.01472}{{\ttfamily arXiv:1809.01472
		[hep-ph]}}.

\bibitem{Bause:2021prv}
R.~Bause, G.~Hiller, T.~H\"ohne, D.~F. Litim, and T.~Steudtner, ``{B-anomalies
	from flavorful U(1)$'$ extensions, safely},''
\href{https://dx.doi.org/10.1140/epjc/s10052-021-09957-1}{{\em Eur. Phys. J.
		C} {\bfseries 82} no.~1, (2022) 42},
\href{https://arxiv.org/abs/2109.06201}{{\ttfamily arXiv:2109.06201
		[hep-ph]}}.

\bibitem{Nadolsky:2008zw}
P.~M. Nadolsky, H.-L. Lai, Q.-H. Cao, J.~Huston, J.~Pumplin, D.~Stump, W.-K.
Tung, and C.~P. Yuan, ``{Implications of CTEQ global analysis for collider
	observables},'' \href{https://dx.doi.org/10.1103/PhysRevD.78.013004}{{\em
		Phys. Rev. D} {\bfseries 78} (2008) 013004},
\href{https://arxiv.org/abs/0802.0007}{{\ttfamily arXiv:0802.0007 [hep-ph]}}.

\bibitem{Rizzo:2014xma}
T.~G. Rizzo, ``{Exploring new gauge bosons at a 100 TeV collider},''
\href{https://dx.doi.org/10.1103/PhysRevD.89.095022}{{\em Phys. Rev. D}
	{\bfseries 89} no.~9, (2014) 095022},
\href{https://arxiv.org/abs/1403.5465}{{\ttfamily arXiv:1403.5465 [hep-ph]}}.

\bibitem{Melnikov:2006kv}
K.~Melnikov and F.~Petriello, ``{Electroweak gauge boson production at hadron
	colliders through $O(\alpha_s^2)$},''
\href{https://dx.doi.org/10.1103/PhysRevD.74.114017}{{\em Phys. Rev. D}
	{\bfseries 74} (2006) 114017},
\href{https://arxiv.org/abs/hep-ph/0609070}{{\ttfamily
		arXiv:hep-ph/0609070}}.

\bibitem{Montano-Dominguez:2022ytt}
J.~Monta\~no Dom\'\i{}nguez, B.~Quezadas-Vivian, F.~Ram\'\i{}rez-Zavaleta, and
E.~S. Tututi, ``{Weak dipole moments of heavy fermions with flavor violation
	induced by Z' gauge bosons},''
\href{https://dx.doi.org/10.1088/1361-6471/ac69ff}{{\em J. Phys. G}
	{\bfseries 49} no.~7, (2022) 075004},
\href{https://arxiv.org/abs/2206.07641}{{\ttfamily arXiv:2206.07641
		[hep-ph]}}.

\bibitem{McEntaggart:2022hey}
A.~McEntaggart, A.~E. Faraggi, and M.~Guzzi, ``{Precision studies for string
	derived $Z'$ dynamics at the LHC},''
\href{https://dx.doi.org/10.1140/epjc/s10052-023-11223-5}{{\em Eur. Phys. J.
		C} {\bfseries 83} no.~1, (2023) 54},
\href{https://arxiv.org/abs/2211.15905}{{\ttfamily arXiv:2211.15905
		[hep-ph]}}.

\bibitem{Gutierrez-Rodriguez:2015qka}
A.~Guti\'errez-Rodr\'\i{}guez and M.~A. Hern\'andez-Ruiz, ``{$Z'$ resonance and
	associated $Zh$ production at future Higgs boson factory: ILC and CLIC},''
\href{https://dx.doi.org/10.1155/2015/593898}{{\em Adv. High Energy Phys.}
	{\bfseries 2015} (2015) 593898},
\href{https://arxiv.org/abs/1506.07575}{{\ttfamily arXiv:1506.07575
		[hep-ph]}}.

\bibitem{Gutierrez-Rodriguez:2018jfl}
A.~Guti\'errez-Rodr\'\i{}guez and M.~A. Hern\'andez-Ru\'\i{}z, ``{Probing the
	Z' resonance and the Higgs boson production and decay in a
	SU(3)$_{L}$\ensuremath{\otimes}U(1)$_{N}$ model at future e$^{+}$e$^{-}$
	colliders},'' \href{https://dx.doi.org/10.1140/epjp/i2018-12202-0}{{\em Eur.
		Phys. J. Plus} {\bfseries 133} no.~9, (2018) 352}.

\bibitem{Ramirez-Sanhez:2018xnh}
F.~Ram\'\i{}rez-S\'anhez, A.~Guti\'errez-Rodr\'\i{}guez,
A.~Gonz\'alez-S\'anchez, and M.~A. Hern\'andez-Ru\'\i{}z, ``{$Z'$ and Higgs
	Boson Production Associated with a Top Quark Pair as a Probe of the
	$U(1)_{(B-L)}$ Model at $e^+ e^-$ Colliders},''
\href{https://dx.doi.org/10.1155/2018/8523854}{{\em Adv. High Energy Phys.}
	{\bfseries 2018} (2018) 8523854}. [Erratum: Adv.High Energy Phys. 2018,
5857287 (2018)].

\bibitem{Allanach:2018lvl}
B.~C. Allanach and J.~Davighi, ``{Third family hypercharge model for $
	{R}_{K^{\left(\ast \right)}} $ and aspects of the fermion mass problem},''
\href{https://dx.doi.org/10.1007/JHEP12(2018)075}{{\em JHEP} {\bfseries 12}
	(2018) 075}, \href{https://arxiv.org/abs/1809.01158}{{\ttfamily
		arXiv:1809.01158 [hep-ph]}}.

\bibitem{Allanach:2021kzj}
B.~C. Allanach, J.~E. Camargo-Molina, and J.~Davighi, ``{Global fits of third
	family hypercharge models to neutral current B-anomalies and electroweak
	precision observables},''
\href{https://dx.doi.org/10.1140/epjc/s10052-021-09377-1}{{\em Eur. Phys. J.
		C} {\bfseries 81} no.~8, (2021) 721},
\href{https://arxiv.org/abs/2103.12056}{{\ttfamily arXiv:2103.12056
		[hep-ph]}}.

\bibitem{Allanach:2020kss}
B.~C. Allanach, ``{$U(1)_{B_3-L_2}$ explanation of the neutral current
	$B$\ensuremath{-}anomalies},''
\href{https://dx.doi.org/10.1140/epjc/s10052-021-08855-w}{{\em Eur. Phys. J.
		C} {\bfseries 81} no.~1, (2021) 56},
\href{https://arxiv.org/abs/2009.02197}{{\ttfamily arXiv:2009.02197
		[hep-ph]}}. [Erratum: Eur.Phys.J.C 81, 321 (2021)].

\end{thebibliography}

\end{document}